# Corrosion fatigue crack initiation in ultrafine-grained near-α titanium alloy PT7M prepared by Rotary Swaging

V.N. Chuvil'deev[a], V.I. Kopylov[a,b], N.N. Berendeev[a], A.A. Murashov[a], A.V. Nokhrin[a(*)], M.Yu. Grayznov[a], I.S. Shadrina[a], N.Yu. Tabachkova[c], C.V. Likhnitskii[a], D.N. Kotkov[a], P.V. Tryaev[d]

[a] Lobachevsky State University of Nizhny Novgorod, 23 Gagarina ave., Nizhny Novgorod, Russian Federation, 603950

[b] Physics and Technology Institute, National Academy of Sciences of Belarus, 10 Kuprevich st., Minsk, Belarus, 220141

[c] National University of Science and Technology "MISIS", 4 Leninskiy ave., Moscow, Russian Federation, 119049

[d] Afrikantov OKBM JSC, Russian Nuclear Corporation "ROSATOM", 15 Burnakovsky proezd, Nizhny Novgorod, Russian Federation, 603074

E-mail: chuvildeev@nifti.unn.ru

---

(*) Corresponding author (nokhrin@nifti.unn.ru)

**Abstract**

The study focuses on corrosion fatigue processes taking place in an ultrafine-grained (UFG) near-α-titanium alloy Ti-2.5Al-2.6Zr (Russian industrial name PT7M) used in nuclear engineering. UFG structure formed with Rotary Swaging is found to increase resistance to corrosion fatigue. Parameters of the Basquin's equation are defined and the slope of the fatigue curve $\sigma_a$-lg(N) is shown to depend (nonmonotonic dependence) on the UFG alloy annealing temperature. This effect can be explained with the patterns of microstructural evolution in a UFG alloy PT7M during annealing: (1) reduced density of lattice dislocations, (2) precipitation and dissolution of zirconium nanoparticles, (3) release of α″-phase particles causing internal stress fields along interphase (α-α″)-boundaries, and (4) intensive grain growth at elevated annealing temperatures. It is shown that the fatigue crack closure effect manifested as changing internal stress fields determined using XRD method may be observed in UFG titanium alloys.

**Keywords:** Titanium alloys; fatigue; fine-grained structure; Rotary Swaging; strength; grain boundary.

**1. Introduction**

Titanium α- and near-α alloys are widely used in nuclear engineering to produce heat exchange equipment for nuclear power plants (NPP) [1, 2]. Titanium alloys for NPPs must comply with most demanding requirements as to strength, ductility, corrosion and radiation resistance [1-5]. In order to meet these requirements, various means are developed to form an ultrafine-grained (UFG) structure with the help of severe plastic deformation (SPD) techniques [6-11].

One of the most dangerous processes damaging heat exchange equipment for NPPs is corrosion fatigue crack initiation [2, 12] that often occurs when metals are subjected to elevated temperatures and hot salt corrosion [12-15]. Despite a keen interest in studying corrosion fatigue processes [12-17], its mechanisms in high-strength UFG titanium alloys are underresearched. (The main attention is paid to the study of fatigue fracture in UFG titanium alloys in the air [18-23]). Therefore, performance predictions for NPPs are ambiguous.

The reasons for corrosive environments having a weak impact on fatigue strength characteristics of titanium alloys are understudied – as shown in [1, 24], fatigue failure curves $\sigma_a$-lg(N) for smooth α- and near-α-alloy specimens during tests in air and in neutral (weak acid) aqueous media are almost coincident, however tests with pre-induced fatigue cracks demonstrate high sensitivity of titanium alloys to corrosion environments [1]. Note references [17, 25] showing that in two-phase titanium alloys (Russian alloys VT6, VT14, etc.) in a neutral environment, fatigue strength characteristics decrease, while in α- and near-α-alloys (Russian alloys PT3V, PT7M) a slightly bigger fatigue limit is observed during tests in a 3% NaCl aqueous solution as compared to fatigue tests in air. Equally contradictory data is provided in literature on microflow contribution to fatigue failure processes in high-ductile titanium alloys, on the impact that grain size and structural-phase state of grain boundaries have on the rate of corrosion fatigue failure in titanium alloys, etc.

It should be noted that most papers currently focus on studying the patterns of corrosion fatigue failure in coarse-grained (CG) ($\alpha+\beta$)-titanium alloys or near-$\beta$ alloys [26-31] that are widely used in aeronautical engineering. Detailed studies related to initiation and propagation of corrosion fatigue cracks in UFG $\alpha$- and near-$\alpha$ titanium alloys extensively used in nuclear power industry are few – expect for [32, 33], as well as studies into corrosion fatigue failure in a UFG pure titanium having various biomedical applications. Such papers [32, 33] as illustrate that formation of an UFG structure increases the fatigue limit, but no detailed studies into the patterns of corrosion fatigue failure were performed.

The research aims to study specific aspects of corrosion fatigue in a high-strength UFG near-$\alpha$-titanium alloy Ti-Al-Zr. The paper focuses mainly on studying the impact of recrystallization annealing on resistance to corrosion fatigue in an UFG near-$\alpha$-titanium alloy Ti-(1.8-2.5)wt.%Al-(2-3)wt.%Zr (Russian industrial name PT7M) that is extensively used today in transport nuclear power industry to produce heat exchange equipment for nuclear marine propulsion secondary systems [2, 3, 5]. This alloy designed for a neutral water-chemical regime of NPP condensate-feeding system is notable for reduced aluminum content, which is conducive to intercrystalline hot salt corrosion when titanium alloys interact with crystalline salts and are exposed to water. Such a situation may occur when salt deposits with elevated concentrations of corrosive components – chlorides and bromides of alkali and alkaline earth metals appear on the surface of heat exchange tubes of steam generators [1-3].

Inadequate strength and corrosion properties of CG titanium alloys simultaneously exposed to mechanical stress and corrosive environments that cause rapid corrosion fatigue and stress corrosion cracking dramatically reduce the life of critical components and structures in modern nuclear marine propulsion plants, increase emergency risks on board the ships, floating nuclear power stations, icebreakers, etc., decrease time between overhauls.

To tackle this challenge arising from increased requirements as to reliability of critical products made from titanium alloys, designers pay special attention to the capabilities of severe plastic deformation ensuring a homogeneous UFG structure in metals – Equal Channel Angular Pressing (ECAP) [11, 34-36], torsion under quasi-hydrostatic pressure (high-pressure torsion, HPT) [7, 37-39], multiaxial forging [6, 40-41], etc.

An UFG structure in a titanium alloy PT7M is formed with Rotary Swaging (RS). RS creates a high-speed impact of bolts on the metal surface where a high-strength UFG structure is formed [42-45]. RS helps efficiently manage the parameters of grain and dislocation (subgrain) metal structure, texture, the degree of nonequilibrium, and local composition of grain boundaries, nature and intensity of internal stresses, etc. This potential explains an ever-increasing interest of researchers in RS capabilities [42-45]. RS, during which a processed workpiece is simultaneously subjected to compression strain and torsion strain, is characterized by complex stress strain behavior that may affect mechanical properties of UFG alloys, including their resistance to initiation and propagation of fatigue cracks.

Note that modern RS machines provide for an opportunity to manufacture long-length pipe workpieces (cylindrical workpieces with a central symmetrical opening), thus opening up greater prospects for RS to be used on a wider scale in order to produce NPP heat exchange equipment from titanium alloys. This paper lays emphasis on analyzing the impact of internal stresses on the tendency of UFG titanium alloys to fatigue failure. Researchers while studying fatigue failure mechanisms currently focus on analyzing the impact that UFG structure parameters and mechanical properties have on fatigue life of UFG alloys [18, 19, 21-23, 32, 34, 36, 43]. The impact of internal stresses that necessarily occur in the metal when a UFG structure is formed with SPD is deemphasized. However, this issue may appear relevant for UFG metals, since the nature (tensile, compressing) and size of internal stress fields formed during severe plastic deformation may significantly affect the growth of fatigue cracks, including through the microcrack closure mechanism [46-50] or changing crack growth rate.

## 2. Materials and methods

The target of research is a near-α industrial alloy PT7M with Ti-2.5wt.%Al-2.6wt.%Zr composition (produced by Chepetsky Mechanical Plant JSC, Glazov, Russia). The concentration of oxygen, nitrogen, hydrogen, and carbon in the alloy is 0.12 wt. %, 0.003 wt. %, 0.001 wt. %, and 0.028 wt.%, respectively. The chemical composition of the alloy is provided inTable1. The chemical composition of the alloy complies with Russian standard GOST 19807-91 requirements.

An FG structure was formed with Rotary Swaging using HF5-4-21 HMP machine (Germany). RS was performed at room temperature by means of gradual deformation of ⌀20 mm rod into ⌀16 mm rod (stage 1), ⌀12 mm rod (stage 2), ⌀10 mm rod (stage 3), ⌀8 mm rod (stage 4), and ⌀6 mm rod (stage 5). The deformation rate was 0.5-1 $s^{-1}$. The process of deformation involved high-speed action of hard-alloy bolts against the surface of a titanium rod with simultaneous axial rotation of the rod.

Structure studies were conducted using Leica IM DM metallographic microscope, Jeol JEM-2100 transmission electron microscope (TEM) with JED-2300 energy dispersive X-ray analyzer and Jeol JSM-6490 scanning electron microscope (SEM) with Oxford Instruments INCA 350 energy dispersive (EDS) microanalyzer. To identify the microstructure (in studies conducted with SEM), electrolytic etching was performed at room temperature in a solution of 75% $H_2SO_4$+15% $HNO_3$+10% HF.

X-ray diffraction (XRD) analysis was performed with DRON-3 diffractometer ($CuK_α$ – radiation with wave length λ = 1.54178 Å, angular range 2Θ = 20-100°, sampling interval 0.02°, holding time at point 0.6 s, studies were performed using the Bragg-Brentano method). Diffraction patterns were analyzed using PhasanX 2.0 software (Powder Diffraction Phase Analysis, ver. 2.03). The scanning rate for overview XRD images was 2 °/min; the rate was 0.2 °/min when analyzing peak broadening. To offset the instrumentation interference $b_i$ from the machine into the broadening of XRD peaks, additional measurements of the dependence of broadening on diffraction angle were taken using a reference specimen Silicon powder 99% 325 mesh. Intrinsic broadening of an XRD peak (hkl) was determined using the formula: $\beta_{hkl} = \sqrt{A_{hkl}^2 - b_{hkl}^2}$, where $A_{hkl}$ – experimentally obtained half-width of XRD peaks. The value of internal stresses $\sigma_{int}$ was calculated with Williamson-Hall method based on the inclination angle of $(\beta_{hkl})^2 \cdot \cos^2\Theta_{hkl} - 16\sin^2\Theta_{hkl}$ dependence [53], where $\Theta_{hkl}$ – angle corresponding to maximum intensity of an XRD peak indexed hkl. Internal stresses have been measured using XRD analysis in ∅3 mm specimens cut out from the central part of a ∅ 6 mm titanium alloy rod (the diameter of studied specimens thus corresponds to the diameter of the operating area of specimens for corrosion fatigue tests). The surface of specimens before studies was mechanically polished to reach the degree of roughness less than 1 μm and subsequently subjected to electrochemical polishing. Electrochemical polishing was used to remove a cold-worked (deformed) layer that occurs during mechanical treatment and distorts the results of internal stress studies using the XRD method.

To study the internal stresses in specimens after corrosion fatigue tests, the end of the destroyed specimen was mechanically polished as deep as possible in order to remove the destroyed surface and obtain a plane parallel surface suitable for studying internal stresses using XRD method. A mechanically polished surface (degree of roughness less than 1 μm) was subjected to electrochemical polishing. On average, the removed layer was 0.1 mm thick.

In order to test mechanical properties, Tinius Olsen H25K-S testing machine was used to perform tensile tests on cylindrical specimens with a working area of 3 mm in diameter (Fig. 1a, b). Pursuant to the requirements of the Russian standard GOST 1497-84, three specimens for each structural condition were tested. Microhardness was measured with HVS-1000 hardness tester under 200 g. Prior to microhardness tests, the surface of specimens was mechanically polished with diamond pastes to reach the degree of roughness approximating 1 μm.

Corrosion fatigue tests of cylindrical specimens with a working area of 3 mm in diameter (Type II as per Russian standard GOST 25.502-79, see Fig. 1c, d) were performed in a 3% NaCl aqueous solution following the cantilevered bending scheme (skewness ratio $R_\sigma = -1$) at 50 Hz. The frequency of 50 Hz has been chosen based on recommendations provided by Afrikantov OKBM JSC as the one corresponding to the frequency range within which vibrations are observed in titanium heat exchange equipment of atomic marine plants. The frequency range of 50 Hz is most common among researchers, which greatly simplifies the task of comparing the obtained results against experimental data provided by other authors. The roughness of a working area in specimens was Rz3.2. Fractographic analysis of fractures in specimens after corrosion fatigue testing was performed using Jeol JSM-6490 scanning electron microscope. Prior to the fractographic analysis, the surface of fractures was subjected to ultrasonic cleaning.

Specimens were subjected to annealing for 30 min in a SNOL furnace followed by cooling in ambient air. Accuracy of temperature control was ± 5 °C. The oxygen concentration after annealing met the requirements of the Russian standard GOST 19807-91.

### 3. Results and discussion

PT7M in the initial state has a CG platelike and needlelike structure with β-phase particles stitched along the boundaries of titanium α and α'-phase. Plates of α'-phase in the initial state are ~5-10 μm thick (Fig. 2a). After RS, a ultrafine-grained-subgrain structure is formed with an average fragment size of ~0.2-0.5 μm (Fig. 2b). EDS results show that the boundaries of UFG alloy fragments are zirconium-rich – local concentration of Zr atoms in the crystal lattice is ~1.6-2.0 wt.%, while the amount of zirconium along fragment boundaries approximates 3.1-3.6 wt.% (Fig. 3). The aluminum concentration in fragment boundaries of an UFG alloy is ~1.4-1.8 wt.%, which is somewhat less than the amount of aluminum in the grain volume (~2.1-2.3 wt.%).

Annealing of an UFG alloy PT7M at 400 °C triggers recovery processes accompanied by declining density of lattice dislocations alongside growing annealing temperature. Zirconium particles uniformly distributed in the crystal lattice and along grain boundaries of an UFG titanium alloy precipitate in the alloy structure. The size of zirconium particles is 20-50 nm (Fig. 4). Structural studies show that after annealing at 500-550 °C, recrystallization processes and dissolution of zirconium particles start simultaneously in an UFG alloy. After annealing at 650 °C, the average grain size varies form 2.8-2.0 μm (surface) to 3.5-4 μm (central part of a rod).

Electron-microscopic analysis shows that platelike particles of the second phase (Fig, 5), presumably α″-phase (Fig 6)[1], appear in the volume fraction of recrystallized grains (Fig. 5). Local internal stress fields appear along interphase (α-α″)-boundaries because of differences in the lattice structure of α-phase and α″-phase (Fig. 5).

After annealing at 600-700 °C, the density of lattice dislocations appears to be very small, no zirconium particles are found in the structure of a recrystallized alloy. Research findings show that the average size of a recrystallized grain on the surface of a titanium rod after annealing at 600-700 °C appears to be 0.5-1 μm less than the average grain size in the center of a cross-section in an UFG titanium alloy PT7M specimen (Fig. 7). Inhomogeneity of the grain structure in a recrystallized specimen is clearly visible in the macrosection showing the area of particular etchability of an UFG structure in the central part of a rod (Fig. 8). Zirconium particles in a recrystallized alloy are absent. The dependence of an average grain size on the annealing temperature in an UFG alloy PT7M is shown in Fig. 9.

For further analysis, it shall be noted that at the initial stage of recrystallization annealing, a heterogeneous consertal structure (see Fig. 10), typical of primary recrystallization stage is formed in a UFG PT7M alloy. It is well-known that primary recrystallization taking place during annealing of severely deformed metals is triggered by a reduction in the reserve internal energy associated with high density of lattice dislocations [51, 52]. As a result, grain boundaries that quickly migrate in case of primary recrystallization actively absorb lattice dislocations and low-angle boundaries formed by them.

---

[1] XRD analysis of a recrystallized alloy PT7M shows no α″-phase – X-ray patterns have no peaks other than α-phase peaks.

Fig. 11 shows XRD findings for near-α-alloy PT7M specimens in various structural states. A CG alloy has all the major peaks corresponding to α'-phase (see Fig. 11a). XRD analysis reveals no β-phase particles in the parent CG alloy, which obviously relates to their small volume fraction. XRD analysis proves that RS causes broadening of α-Ti peaks and their slight shift towards the area of small reflection angles – diffraction angle $2\Theta_{max}$ for α-Ti XRD maximum (100) in an UFG alloy PT7M after RS increases from 35.11° to 35.05-35.07°, while its half-width at half-height ($\beta_{0.5}$) goes up from 0.308° to 0.554° (Fig. 11b) A shift of XRD peaks towards the area of small diffraction angles in line with Wulff–Bragg's condition ($2d_{hkl} \cdot \sin\Theta_{max} = n\lambda$, where $d_{hkl}$ – interplanar spacing) indicates a decrease in interplanar spacing $d_{hkl}$ and formation of compressive internal stresses ($\sigma_{int}$) after RS. The stress value $\sigma_{int}$ determined with Williamson-Hall method [53] by the inclination angle of $(\beta_{0.5})^2 \cdot \cos^2\Theta_{max} - 16\sin^2\Theta_{max}$ dependence (Fig. 12a) amounts to 160 ± 20 MPa for a CG alloy and 520 ± 40 MPa for an UFG alloy PT7M after RS. XRD analysis results show that annealing at 300 °C, 400 °C, and 500 °C brings internal stresses ($\sigma_{int}$) down to 420 ± 30 MPa, 330 ± 30 MPa, and 170 ± 20 MPa, respectively. A further increase in the annealing temperature to 700 °C slightly reduces $\sigma_{int}$ to 160 ± 20 MPa (Fig. 12b).

Fig. 13a shows stress-strain dependences for UFG alloy PT7M specimens after different annealings, while Fig. 13b shows dependences of yield stress, ultimate tensile strength, and ultimate elongation to failure on the temperature during 30 min annealing of an UFG alloy PT7M. Tensile tests show that formation of an UFG structure leads to an increase in yield stress $\sigma_{0.2}$ and ultimate tensile strength $\sigma_b$ from 280 to 1070 MPa and from 590 to 1080 MPa, respectively. Ultimate elongation decreases from 40% to 6%. Studies show that an UFG alloy in the cross-section loses its hardness from the edge to the center of a rod, hardness falling from 2.9-3.0 to 2.5-2.6 GPa. After annealing at 700 °C (30 min), the difference in hardness between the central and surface areas of the cross-section of a rod does not exceed 0.1-0.15 GPa. Mechanical properties of annealed PT7M: $\sigma_{0.2}$=350-370 MPa, $\sigma_b$=600-620 MPa, $\delta_5$=12-14%.

Fig. 14 shows fatigue curves $\sigma_a$-lg(N) for a CG alloy (Fig. 14a) and an UFG alloy (Fig. 14b). Fig. 11 also shows fatigue curves corresponding to 50% probability of no-failure operation. The index of dispersion SI and the reverse angle of inclination $k_1$ are indicated for each of the curves in the Table 2. The conventional endurance limit $\sigma_{-1}$ based on $N^* = 3 \cdot 10^6$ cycles for a CG alloy is 320-330 MPa, while for an UFG alloy after RS $\sigma_{-1}$ = 570-590 MPa. Note that the said increase in the fatigue life of a UFG material occurs when the material plasticity decreases (relative elongation to break goes down from 40% for a CG structure to 7% for a UFG structure) and the typical grain size shrinks. After annealing at 650 °C, $\sigma_{-1}$ goes down to ~500 MPa. Note that formation of an UFG structure increases the durability ($N_f$) of an alloy at $\sigma_a$ = 600 MPa from $10^3$ to $4 \cdot 10^4$-$5 \cdot 10^5$ stress cycles. An increase in annealing temperature from 500 to 650 °C at $\sigma_a$ = 600 MPa leads to a decline in the number of load cycles $N_f$ from $7 \cdot 10^5$ to $1.5 \cdot 10^5$. Fatigue curves $\sigma_a$-lg(N) for a fine-grained PT7M alloy after recrystallization annealing are presented in Fig. 14c.

Note that with an UFG alloy PT7M, a wider scatter of data is observed as compared to specimens of an alloy in an initial state (Fig. 14a, b). We reckon that it is due to inhomogeneity of deformation distribution in the cross-section of a rod (Fig. 7). This assumption is indirectly confirmed by reduced test point scattering in $\sigma_a$-lg(N) dependence when the temperature of annealing an UFG alloy PT7M rises, thus causing the formation of a homogeneous grain structure.

Fig. 15 shows the results of fractographic analysis of fractures in CG alloy PT7M specimens. For the sake of convenience, typical areas of a fatigue crack are presented in Fig. 15. The analysis of obtained results proves that the decay of PT7M alloy at high load amplitudes is mostly brittle – the fracture shows numerous transgranular cleavage facets, traces of a transgranular fracture are observed along crystallographic planes (in Fig. 14 these elements are marked CP and are found close to a fatigue crack nucleus). The areas of beachmarks are observed for some favorably oriented elongated grains (such beach marks in. Fig. 14 are marked FB); the distance between beach marks is on average 1 μm. It is noteworthy that the break zone is difficult to discern in fatigue fractures of a CG alloy PT7M because its area is rather small at all amplitudes studied.

Fractures in specimens of an UFG alloy PT7M (Fig. 16) depending on the degree of applied stress range have one or several microcrack initiation nuclei (Zone I), stable crack growth zone (Zone IIa), accelerated crack growth zone (Zone IIb), and break zone (Zone III). The area of Zone III is growing along with $\sigma_a$ and becomes comparable with the area of Zone II at $\sigma_a$ over 700 MPa. Research findings (Fig. 17) show that decay of an UFG alloy PT7M at the stage of stable crack growth is characterized by brittle-ductile failure, typical transgranular cleavage facets are not observed in Zone II due to small grain size in the fracture (see classification [54]). Stable crack growth zone segues to enhanced crack growth zone with secondary cracks (in figures marked SC). This fact indicates high crack propagation rates, which can possibly be explained by a high stress intensity factor. According to [54], secondary cracks reveal ductile failure patterns in the enhanced crack growth area. According to the classification provided in [54], Zone III is characterized by ductile failure and shows multiple pits that apparently form because of merging micropores. Note that the area of stable crack growth does not exceed 40% of the fracture area at all degrees of stress range. This value is significantly less than in case of a CG alloy.

Fractographic analysis shows that with annealing temperature rising to 650 ºC (at specified load amplitude), the number of crack initiation nuclei growths, the area of stable crack growth reduces from 40-45% to 25-30%, the fracture area corresponding to the enhanced crack growth zone increases from 30-35% to 55%. The break zone area during annealing slightly decreases from 25-30% to 15-20%. Comparison of fractures in CG and UFG specimens reveals that the distance between beach marks in CG specimens appears to be much bigger than that in UFG fractures – with resolution being the same, the distance between beach marks in UFG specimens is almost unobservable (Fig. 16), including during fractographic analysis of fractures in UFG specimens after recrystallization annealing (Fig. 17). According to [46, 54], the obtained results indirectly prove that the crack propagation rate in UFG alloys appears to be much less than in CG alloys.

Thus, it can be concluded that formation of an UFG structure in PT7M alloy with RS improves fatigue strength 1.7-1.8 times without altering the failure mechanism.

The least squares method was used for each curve $\sigma_a$-lg(N) to identify the parameters of the Basquin's power-law equation: $\sigma_a = A \cdot N^{-q}$ [55, 56], where A and q are numerical coefficients (see Table 2). It is notable that the dependence of the slope of the fatigue curve $\sigma_a$-lg(N) on the annealing temperature is nonmonotonic – formation of an UFG structure increases parameter A from 750 MPa to 2806-3244 MPa. Annealing at 500 °C reduces A to 1480 MPa, while recrystallization starting at 550-600 °C increases A to 4810 MPa again. After annealing at 650 °C, parameter A goes down to 1570 MPa again (see Table 2). This is rather unexpected because it is generally assumed that while annealing an UFG titanium, the slope of the fatigue curve $\sigma_a$-lg(N) decreases monotonously [57].

A model of fatigue failure proposed in [58] suggests that parameter A is proportionate to the free energy ($\Delta F$) of lattice dislocations overcoming obstacles in the crack tip plastic zone. (Impact of the corrosion environment on fatigue failure in $\alpha$- and near-$\alpha$-titanium alloys is small [17, 24-25] because of fast repair of an oxide film.) According to [59], $\Delta F$ is strongly affected by the nature of obstacles – the degree of lattice distortion caused by atoms of doping elements, size of the second-phase particles, density of lattice dislocations, etc. Changes in the types of obstacles determining the rate of lattice dislocations alter $\Delta F$ and consequently the slope of $\sigma_a$-lg(N) dependence.

A decrease in parameter A after annealing at 400 °C can be explained by reduced density of lattice dislocations as a result of recovery processes that kick in. Formation of $\alpha''$-phase particles obstructing dislocations and formation of internal stress fields along interphase ($\alpha$-$\alpha''$)-boundaries leads to an increase in $\Delta F$. Simultaneous dissolution of zirconium particles blocking the movement of lattice dislocations and intensive grain growth at rising annealing temperatures are expected to reduce $\Delta F$ and consequently the slope of fatigue curve $\sigma_a$-lg(N).

Complicated behavior of internal stresses during annealing UFG titanium alloys can have additive impact on nonmonotonic character of A(T) dependence.

References [60, 61] show that long-range fields of internal stresses have a significant effect on the movement of lattice dislocations in UFG metals. The internal stress rate ($\sigma_{int} = \alpha_1 \rho_b \Delta b + \alpha_2 w_t$) is proportionate to the density of defects in grain boundaries – orientational misfit dislocations with density $\rho_b \Delta b$ and sliding components of delocalized dislocations with density $w_t$ ($\alpha_1$, $\alpha_2$ – coefficients). The density of defects ($\rho_b \Delta b$, $w_t$) in grain boundaries depends on the annealing temperature of UFG alloys, and during recovery annealing, the density of defects in grain boundaries decreases leading to reduced internal stresses $\sigma_{int}$ [60-62].

During recrystallization annealing of an UFG metal, grain boundaries migrating at the rate of $V_m$ bring forth dislocations from the crystal lattice. This results in a flow of lattice dislocations $I = \xi \rho_v V_m$ ($\xi$ - numerical coefficient, $\rho_v$ – density of lattice dislocations) [60-62], increased density of defects in grain boundaries of an UFG alloy, and consequent increase in $\sigma_{int}$. If the annealing temperature continues to rise, $\sigma_{int}$ will fall due to a more intensive diffusion defect recovery in grain boundaries and reduced density $\rho_v$.

We reckon that nonmonotonic changes in $\sigma_{int}$ during annealing of an UFG alloy lead to nonmonotonic changes in the free energy ($\Delta F$) of lattice dislocations overcoming obstacles and therefore to nonmonotonic changes in the slope of fatigue curve $\sigma_a$ – $lg(N)$ during annealing of an UFG alloy PT7M.

The final part hereof discusses the fatigue crack growth mechanism.

Let us proceed with the above phenomenon of changing durability when obtaining a UFG structure with RS. Fractographic analysis of fractures proves that UFG specimens obtained with RS have a smaller area of stable crack growth as compared to CG specimens. Note that according to [46, 47-49, 63], the crack growth rate shall increase against the drop in the grain size. Our observations show that the crack growth rate slows down to the extent that beach marks for UFG specimens are unobservable (Fig. 15).

Note that according to XRD data, significant tensile residual stresses (+520 MPa) occur after RS in a titanium alloy that shall accelerate crack growth and reduce fatigue life as compared to UFG titanium and near-α titanium alloys obtained with other methods of severe plastic deformation that would not cause bigger internal stresses (for example, see [18, 19, 23, 36, 57]). When the results of fatigue[2] studies of a UFG near-α PT7M titanium alloy obtained with RS are compared in this work (see also [43]) with the results of fatigue studies of a commercial-purity UFG titanium alloy obtained with ECAP [18, 19, 21, 32, 34, 36, 57], multiaxial forging [41, 64], and hydrostatic extrusion [42], it turns out that with approximately the same grain size (~ 0.5 μm), the fatigue life of a UFG titanium after ECAP or multiaxial forging appears to be much higher.

This effect proves that during fatigue tests of a UFG PT7M alloy, there should be a factor that would reduce the crack growth rate. Under the conditions studied, according to [48, 57, 59, 60], crack closure caused for instance by plastic deformation might be such a factor. This effect is associated with residual compression stresses that occur in the plasticity area along crack edges. It is generally assumed that the value of internal stresses in the plastic deformation area corresponds to the yield stress of the material, while the plastic deformation area depends on the stress range [48, 58]

It can be expected that the negative effect of tensile internal stresses that occur during RS is compensated by the formation of compressing fields of internal stressesat the top of the crack and along the edges of a growing fatigue crack. This assumption is indirectly corroborated by the results of XRD studies of internal stresses (Fig. 18, 19).

Studies show that the subsurface fracture area in CG specimens creates compressing fields of internal stresses, an X-ray peaks α-Ti shifts towards bigger diffraction angles $2\Theta_{max}$ (Fig. 18). Residual internal stresses in a CG PT7M alloy grow together with a test range (Fig. 19a). A similar analysis of a subsurface area of fractures in UFG alloys shows that tensile internal stresses reduce in UFG alloys $\sigma_{int}$. This effect is particularly evident within large stress ranges (Fig. 19). We reckon that the obtained results indirectly confirm the feasibility of a plastic crack closure mechanism in PT7M Thus, it might be assumed that high fatigue characteristics of a UFG PT7M alloy are mostly due to a high degree of resistance to nucleation of fatigue microcracks that depends on deformation defects present along grain boundaries of a UFG alloy, as well as internal stress fields along interphase ($\alpha$-$\alpha''$)-boundaries when $\alpha''$- phase particles form in recrystallized alloys. The difference in the crack growth rate in CG and UFG alloys manifested as a changing distance between beach marks in the fractures is less pronounced due to a negative impact of tensile internal stress fields that appear during RS.

---

[2] In case of titanium, the impact of the corrosive environment on the fatigue test results can be neglected due to its high corrosion resistance.

**Conclusion**

1. Rotary Swaging was used to obtain specimens of an UFG titanium near-$\alpha$-alloy PT7M characterized by high hardness, high yield stress, and high ultimate tensile strength. UFG PT7M specimens obtained with RS are characterized by significant (+520 MPa) tensile internal stresses. Formation of an UFG structure is shown to increase conventional endurance limit $\sigma_{-1}$ (based on $N^* = 3 \cdot 10^6$ test cycles) from 320-330 MPa to 570-590 MPa and durability $N_f$ (at load amplitude $\sigma_a$ = 600 MPa) from $10^3$ stress cycles to $4 \cdot 10^4$-$5 \cdot 10^5$ stress cycles. Corrosion fatigue tests of UFG alloy specimens are found to be characterized by a wide test point scattering caused by inhomogeneous deformation distribution along the longitudinal section of a titanium rod after RS.

2. The corrosion fatigue curve $\sigma_a$ – lg(N) for an UFG alloy PT7M is shown to be accurately described with the Basquin's equation. Formation of an UFG structure in PT7M alloy is found to increase the slope of $\sigma_a$ – lg(N) dependence in the low-cycle fatigue area (parameter A in the Basquin's equation rises from 750 MPa to 2806-3244 MPa). Parameter A in the Basquin's equation is found to be in nonmonotonic (maximum) dependence on the annealing temperature in an UFG alloy PT7M. A qualitative explanation for nonmonotonic A(T) dependence is proposed.

3. It is shown that the nature of A(T) dependence can be explained with evolution patterns in the microstructure of a highly deformed PT7M alloy during annealing: (1) reduced density of lattice dislocations, (2) precipitation and dissolution of zirconium nanoparticles, (3) precipitation of $\alpha''$-phase particles leading to internal stress fields along interphase ($\alpha$-$\alpha''$)-boundaries, and (4) intensive grain growth at elevated annealing temperatures. XRD method proves that after fatigue tests, the subsurface fracture layer forms compressive internal stresses. It may testify to the mechanism of plastic crack closure. Thus, the negative impact of tensile internal stresses that occurs when a UFG structure is formed with RS can be compensated.

**Acknowledgment**

The research was supported by the Russian Science Foundation (Grant No. 16-13-00066).

**Data Availability**

The raw/processed data required to reproduce these findings cannot be shared at this time as the data also forms part of an ongoing study.

Table 1. Chemical composition of PT7M alloy(in wt.%)

| Ti | Al | Zr | V | Si | Fe | Sn | Nb | $O_2$ | $H_2$ | $N_2$ | C |
|---|---|---|---|---|---|---|---|---|---|---|---|
| balance | 2.45 | 2.63 | 0.002 | 0.02 | 0.086 | 0.005 | 0.024 | 0.12 | 0.001 | 0.003 | 0.028 |

Table 2. Values of constants in the Basquin's equation for a titanium alloy PT7M in various structural states

| Material state | Basquin's equation parameters | | Inverse slope of the $\sigma$-lgN plots | Scatter-index of the $\sigma$-N plots |
|---|---|---|---|---|
| | A, MPa | q | | |
| Initial state | 750 | 0.06 | 16.1 | 0.07 |
| UFG alloy after RS | 2806-3244 | 0.12-0.14 | 6.7-8.3 | 0.14-0.16 |
| RS + annealing at 500 °C | 1480 | 0.07 | 14.3 | 0.10 |
| RS + annealing at 550 °C | 4810 | 0.20 | 4.8 | 0.14 |
| RS + annealing at 600 °C | 3545 | 0.21 | 5.0 | 0.05 |
| RS + annealing at 650 °C | 1570 | 0.10 | 10.0 | 0.17 |

**List of figures**

**Figure 1**. Drawing (a, c) and photos (b, d) of a specimen for tensile tests (a, b) and corrosion fatigue tests (c, d)

**Figure 2.** Structure of PT7M alloy: a) microstructure of an alloy in the initial coarse-grained state; b-d) microstructure of an UFG alloy after RS

**Figure 3.** EDS analysis of the grain boundaries composition in UFG alloy Ti-2.5Al-2.6Zr after RS

**Figure 4.** Microstructure of PT7M alloy after RS and annealing. Zirconium particles in the structure of PT7M alloy after annealing at 400 °C (a, b), 600 °C (c, d) and EDS analysis of their composition

**Figure 5.** α″-phase particles in recrystallized α-titanium grains after annealing at 500 °C (a), 600 °C (b) and 700 °C (c). Fig.5a, 5c, 5d show stress fields along interphase boundaries (sites where α″-phase particles yield to α-phase particles) marked with black arrows, α″-phase particles are marked with white arrows

**Figure 6**. TEM-analysis of the α″-phase particles in recrystallized α-titanium grains after annealing at 500 °C (see Fig.5a). In Fig.6a c – diffraction pattern. In Fig. 6b, c – dark-field image in α-phase (Fig. 6b) and α″-phase (Fig. 6c) reflex

**Figure 7.** Microstructure of a surface (a) and central (b) layers of PT7M titanium alloy rod after RS and annealing at 650 °C (30 min)

**Figure 8.** Macrostructure of a cross-section (a) and longitudinal section (b) in specimens of titanium alloy PT7M after RS and recrystallization annealing at 650 °C (30 min)

**Figure 9.** Dependence of the average grain size on the annealing temperature during 30 min annealing of an UFG alloy PT7M: (1) center of a specimen; (2) edges (surface) of a specimen

**Figure 10**. Microstructure of a central layers of PT7M titanium alloy rod after RS and annealing at 600 °C (30 min)

**Figure 11.** XRD analysis of PT7M alloy in the initial state (a) and in an UFG state after RS (b)

**Figure 12.** XRD analysis: a) Williamson-Hall curves plotted after analyzing XRD patterns of Ti-2.5Al-2.6Zr alloy specimens in the original state (1), in an UFG state after RS (2), after annealing for 400 ºC (3) and for 600 ºC (4); b) dependence of internal stresses on the annealing temperature $\sigma_{int}(T)$ in an UFG alloy PT7M

**Figure 13.** Impact of annealing on mechanical properties of an alloy PT7M: a) strain-stress diagrams: line (1) – CG alloy; line (2) – UFG alloy after RS; line (3) - RS + annealing at 600 ºC; line (4) - RS + annealing at 650 ºC; line (5) - RS + annealing at 700 ºC; b) dependence of mechanical properties of an UFG alloy PT7M on the annealing temperature during 30 min annealing

**Figure 14.** Corrosion fatigue failure curves for PT7M alloy in the initial state (a), in an UFG state after RS (b), and for a fine-grained PT7M alloy after recrystallization annealing at different temperatures (c). Fig. 14a, b also shows fatigue curves corresponding to 50% probability of no-failure operation (dotted line). Fig.14b shows fatigue curves to 50% probability of no-failure operation with Basquin's equation parameters A = 2806 MPa, q = 0.14 (right branch) and A = 3244 MPa, q = 0.12 (left branch).

**Figure 15.** Fractographic analysis of fractures in CG alloy PT7M specimens after corrosion fatigue testing. Stress amplitude $\sigma_a$ = 335 MPa (a), 380 MPa (b), 415 MPa (c). In Fig.15: (1) fracture overall view; (2) fractographic analysis of fractures in Zone I, (3) fractographic analysis of fractures in Zone II, (4) fractographic analysis of fractures in Zone III.

**Figure 16.** Fractographic analysis of fractures in UFG alloy PT7M specimens after corrosion fatigue testing at the amplitude of 575 MPa (a), 905 MPa (b), and 925 MPa (c). In Fig.16: (1) fracture overall view; (2) fractographic analysis of fractures in Zone I, (3) fractographic analysis of fractures in Zone IIa; (4) fractographic analysis of fractures in Zone IIb; (5) fractographic analysis of fractures in Zone III.

**Figure 17.** Fractographic analysis of fractures zones in an UFG alloy PT7M after annealing at 650 ºC. Stress amplitude – $\sigma_a$ = 615 MPa (a), 730 MPa (b), and 925 MPa (c). In Fig.17: (1) fracture overall view; (2) fractographic analysis of fractures in Zone I, (3) fractographic analysis of fractures

in Zone IIa; (4) fractographic analysis of fractures in Zone IIb; (5) fractographic analysis of fractures in Zone III.

**Figure 18.** Results of XRD analysis of the subsurface fracture area in a CG alloy. Comparison by broadening of XRD peak (002) and (101) $\alpha$-Ti (a), and (104) $\alpha$-Ti (b) in CG alloy PT7M at stress ranges $\sigma_a$ = 440 MPa

**Figure 19**. Analysis of the XRD results: a - analysis of results presented in Fig.18 using the Williamson-Hall method (Williamson-Hall curves plotted after analyzing XRD patterns in CG alloy at different stress ranges $\sigma_a$); b - dependence of internal stresses in the subsurface fracture layer of a CG and UFG alloy on the stress range

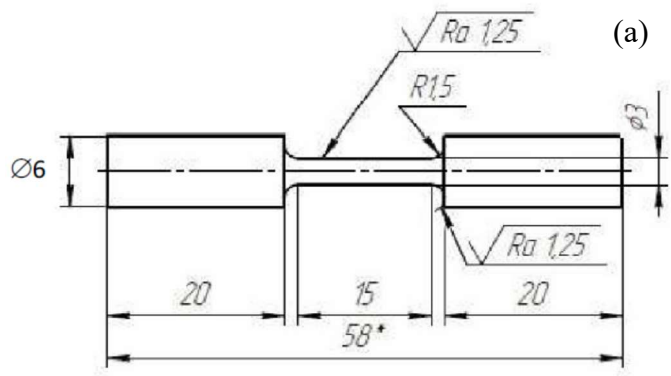 (a)
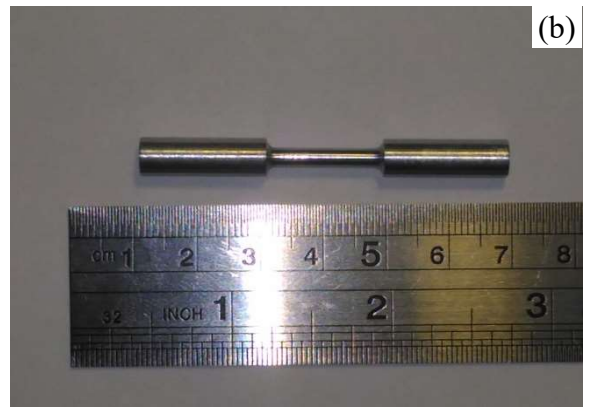 (b)
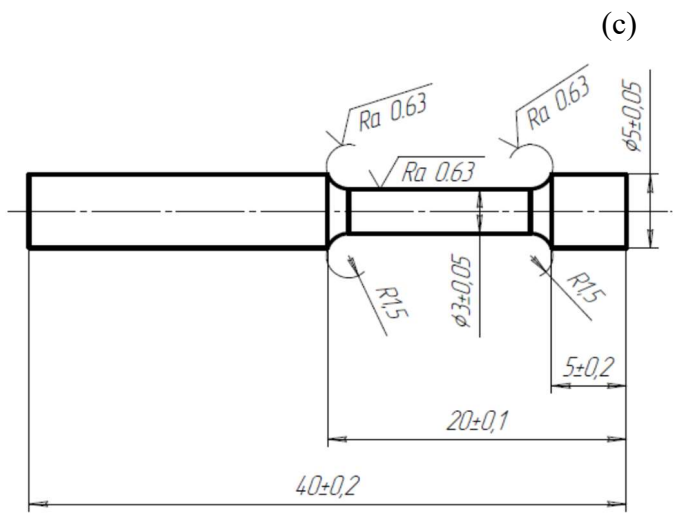 (c)
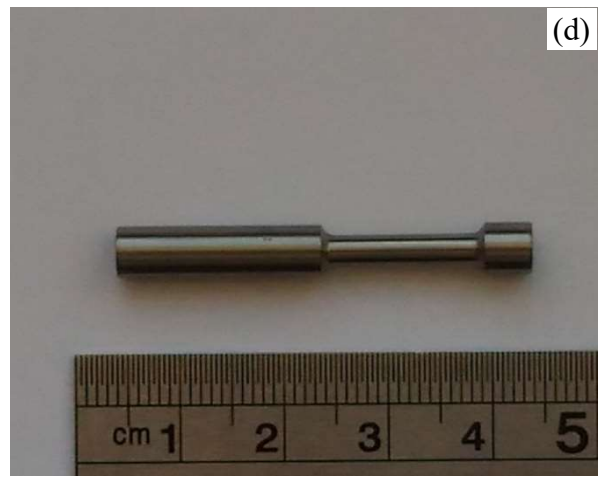 (d)

**Figure 1**

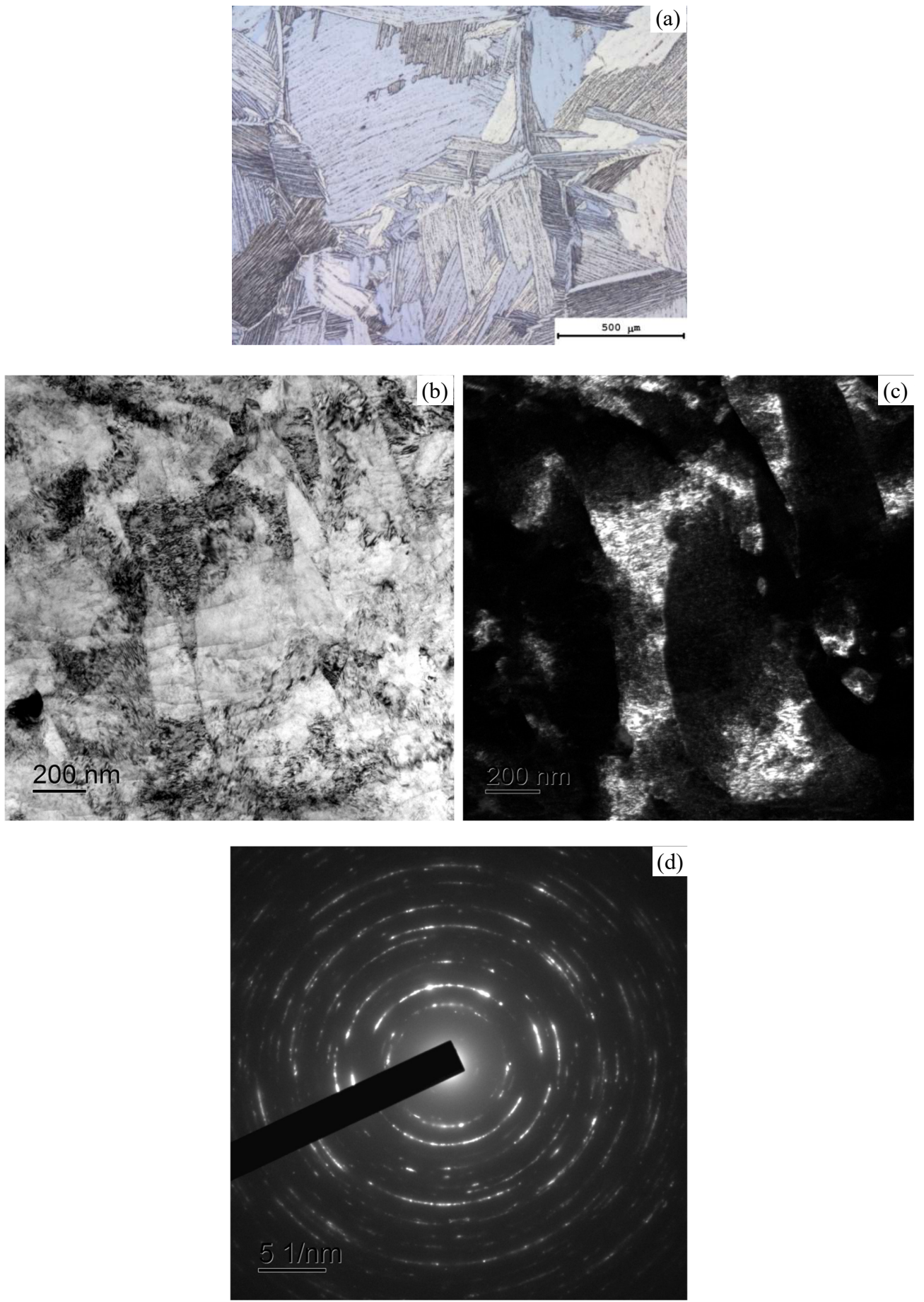

**Figure 2**

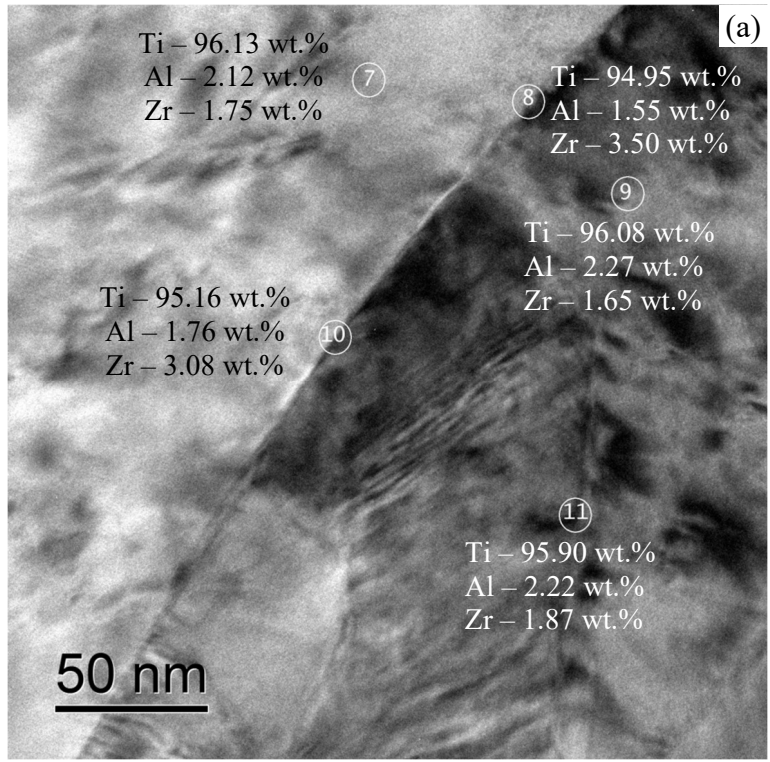

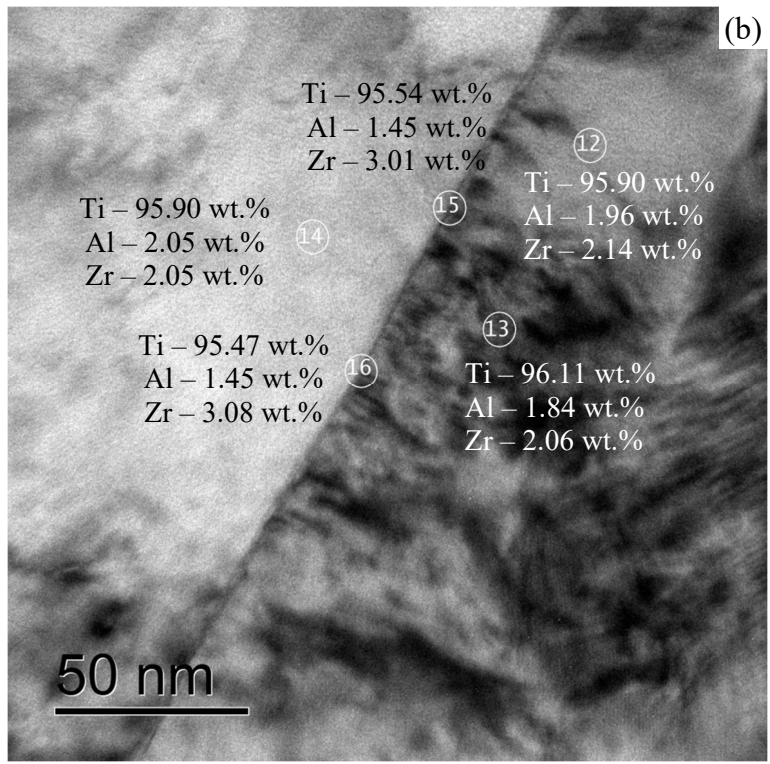

**Figure 3**

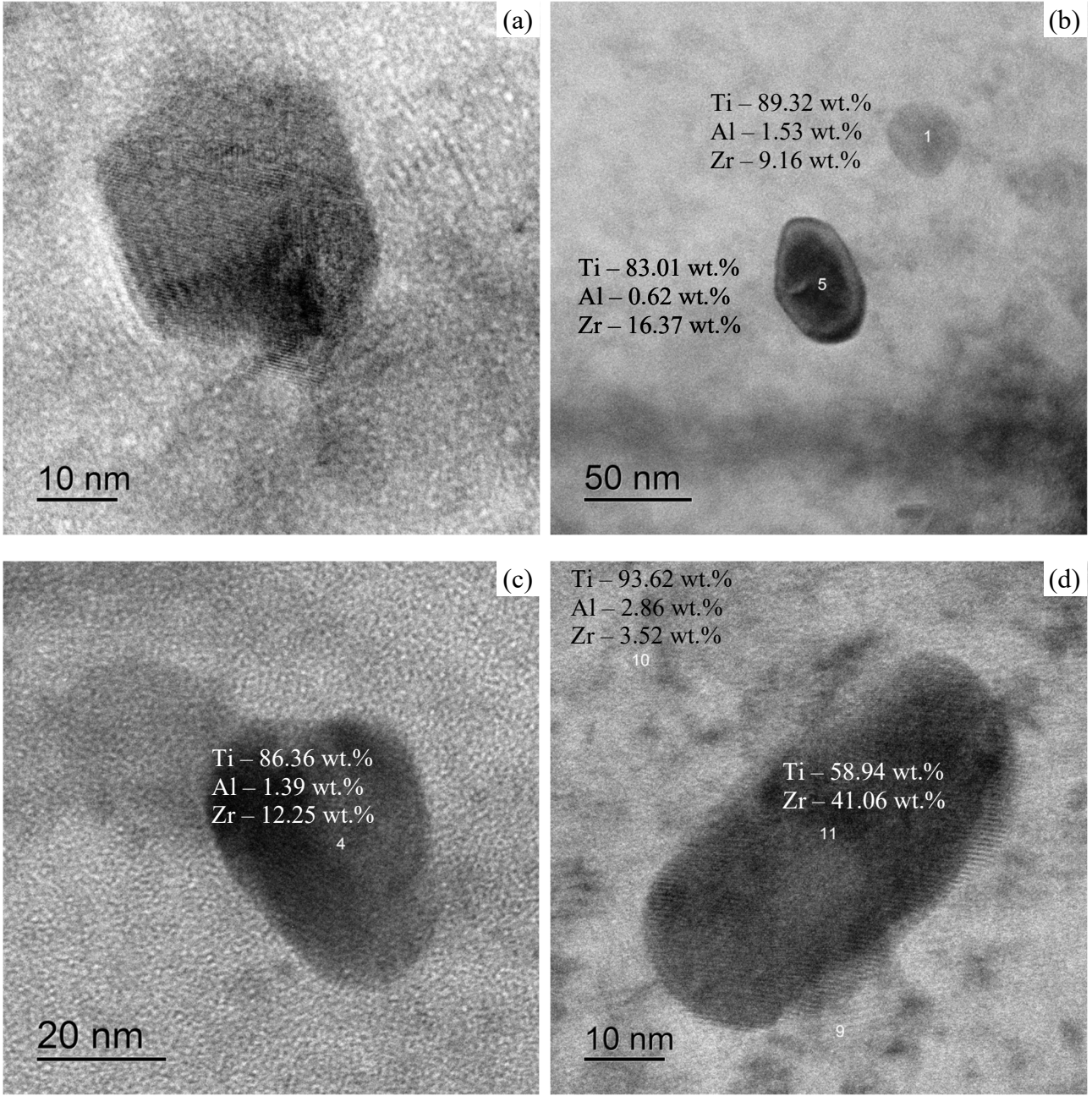

**Figure 4**

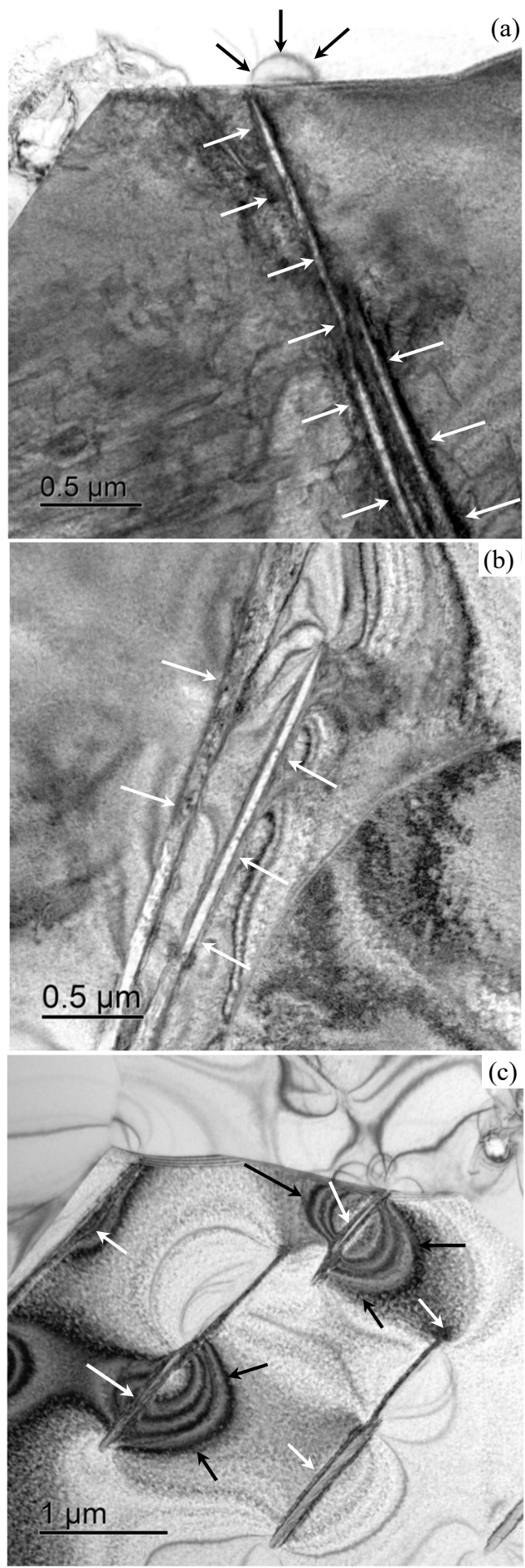

**Figure 5**

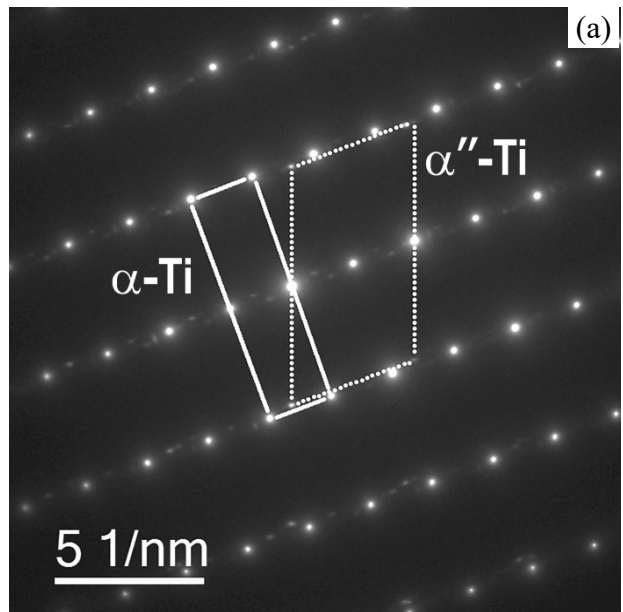

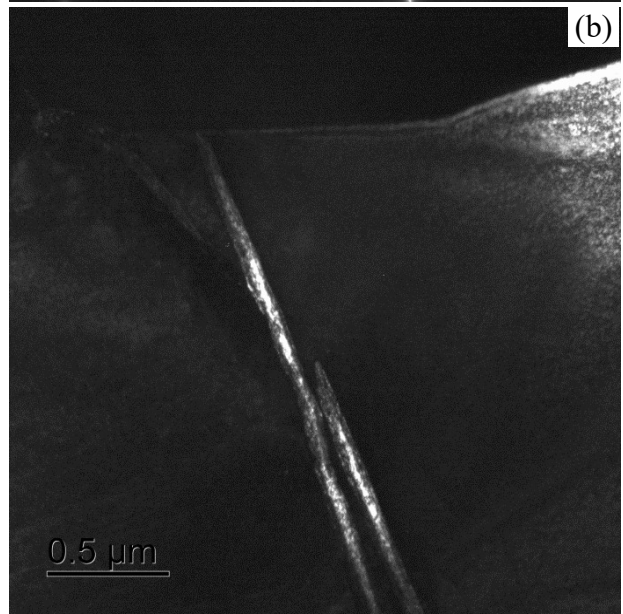

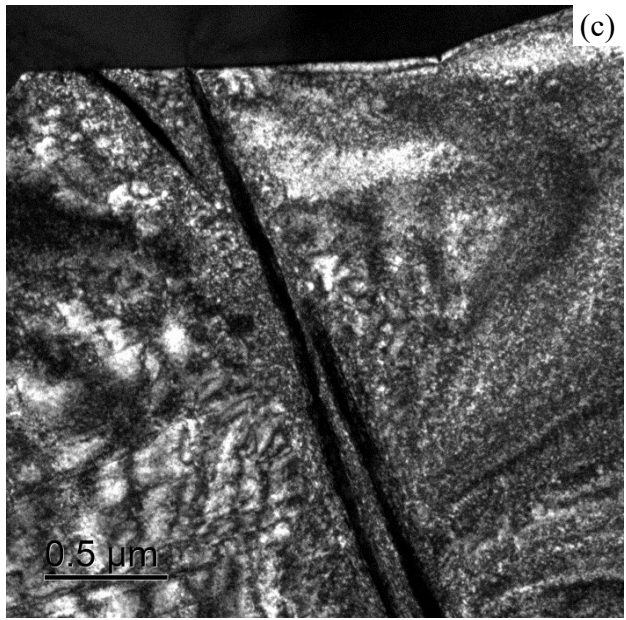

**Figure 6**

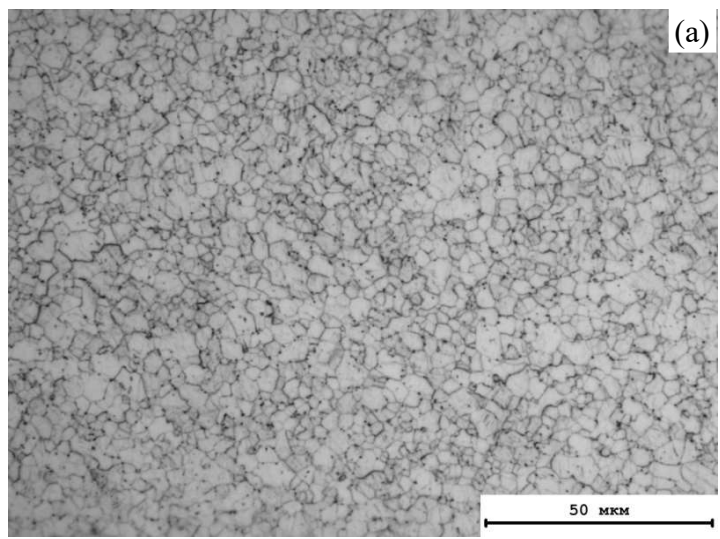

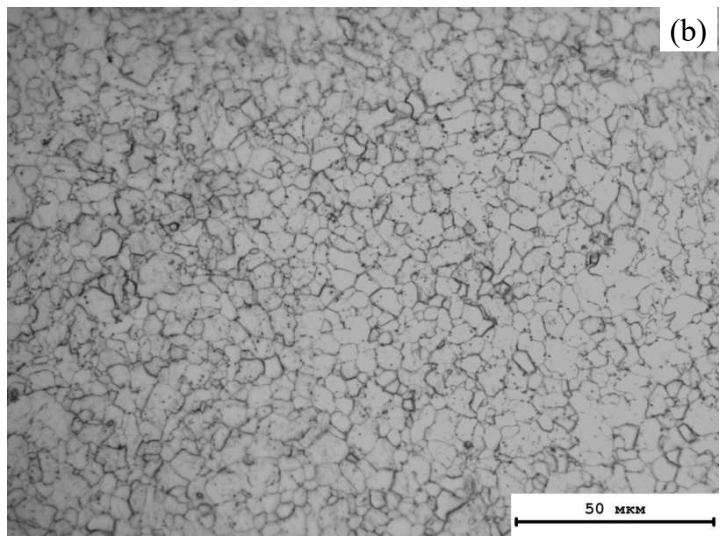

**Figure 7**

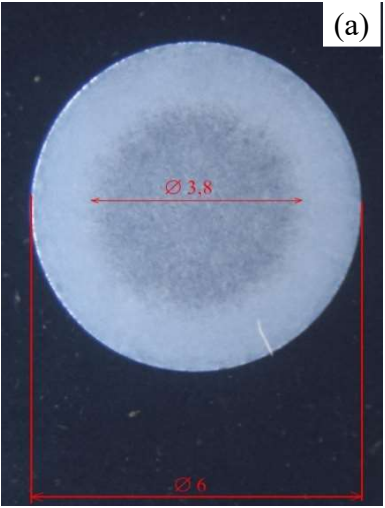

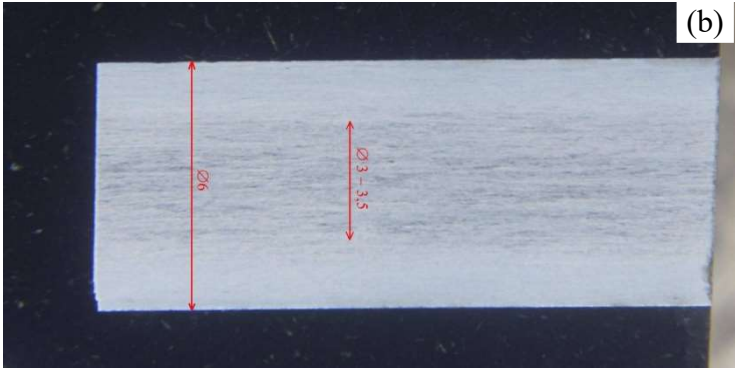

**Figure 8**

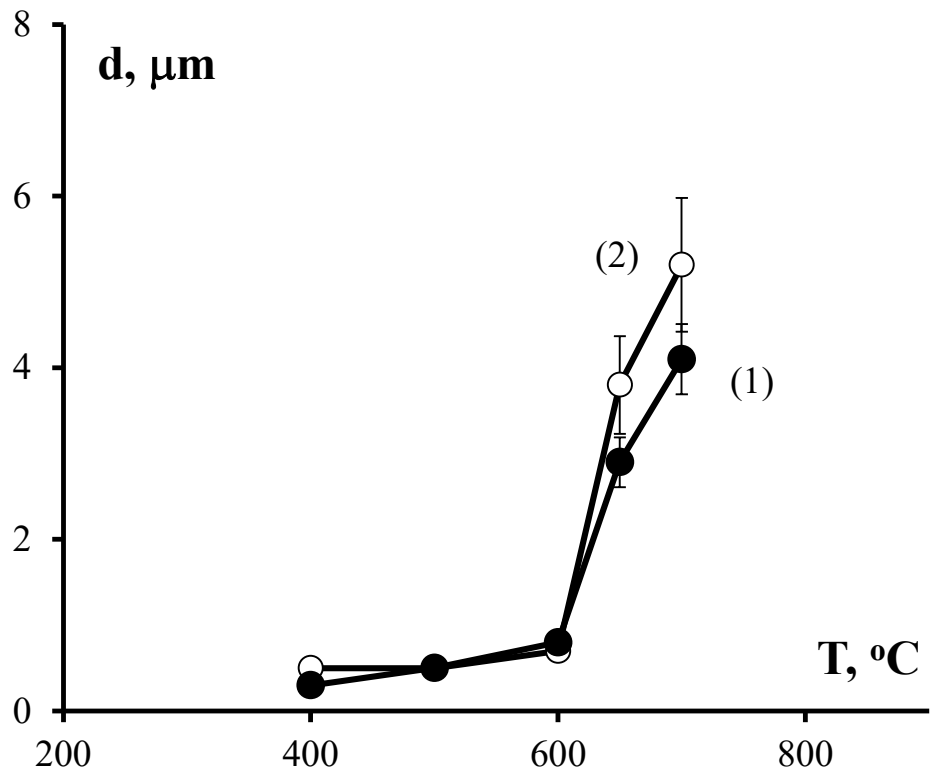

**Figure 9**

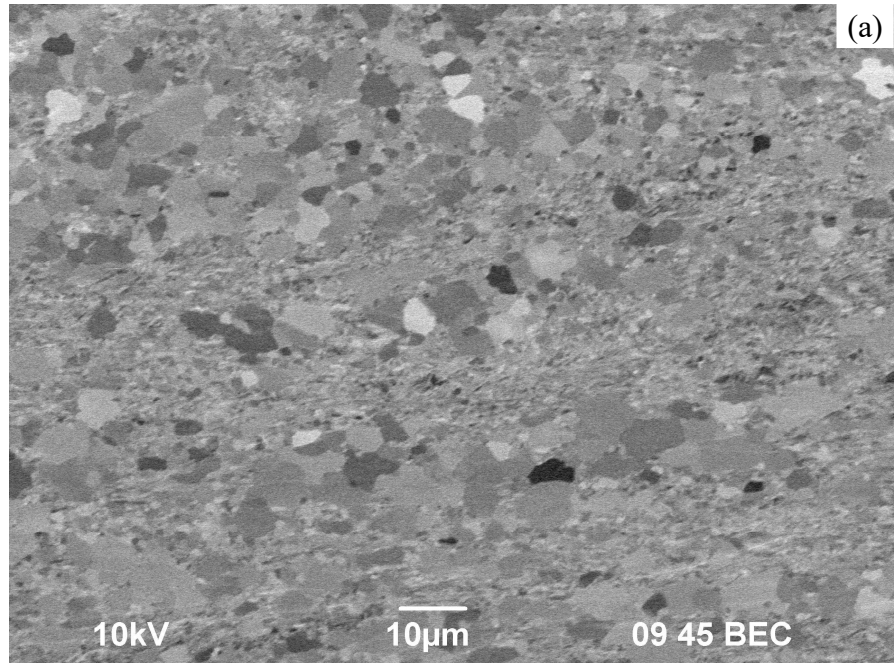

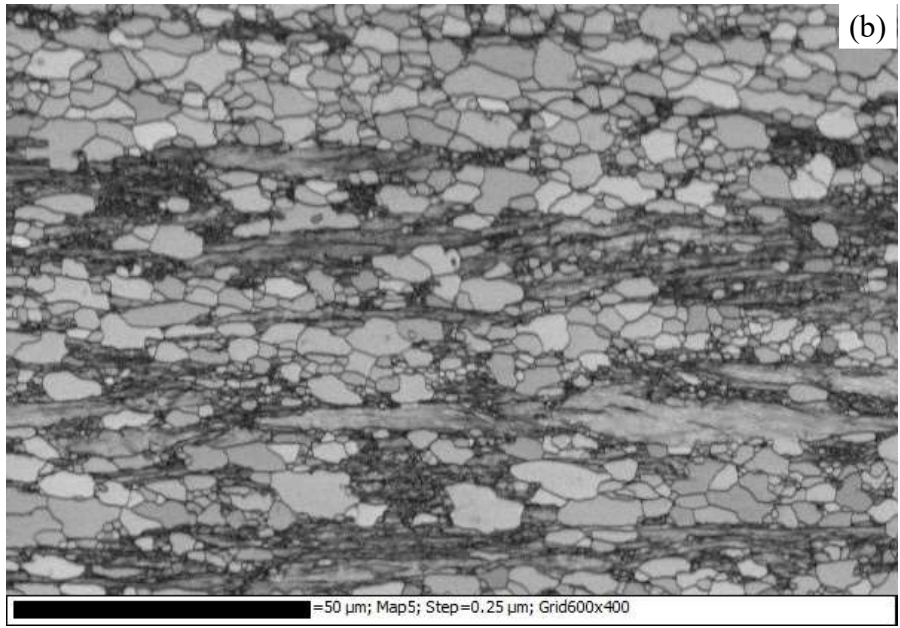

**Figure 10**

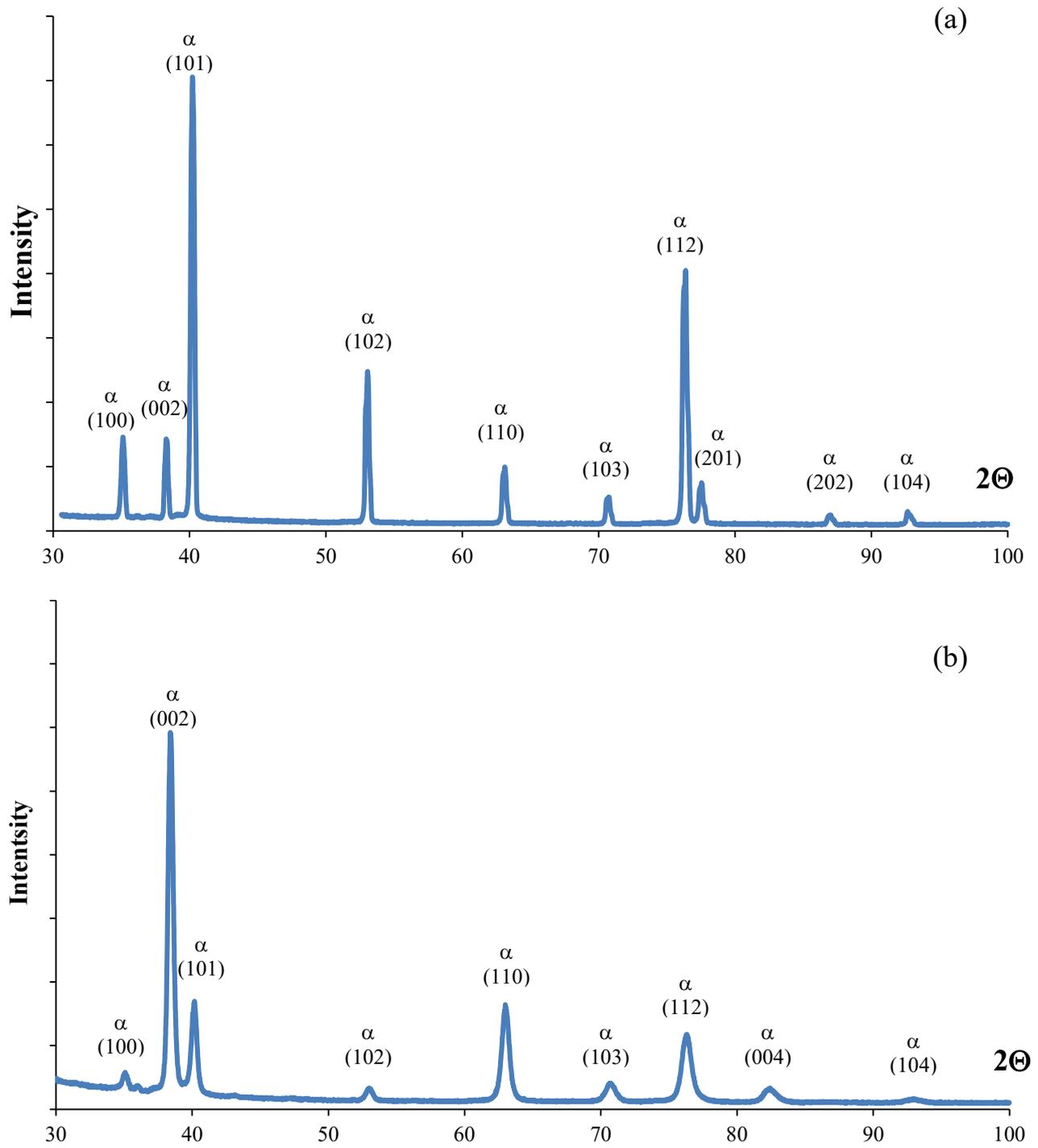

**Figure 11**

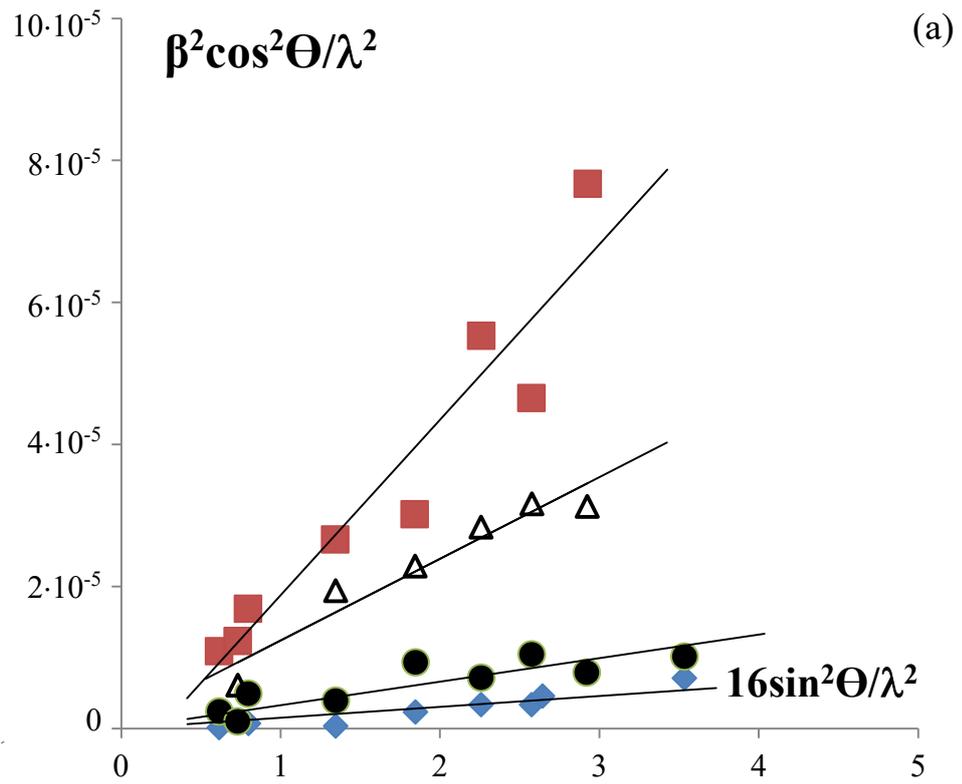

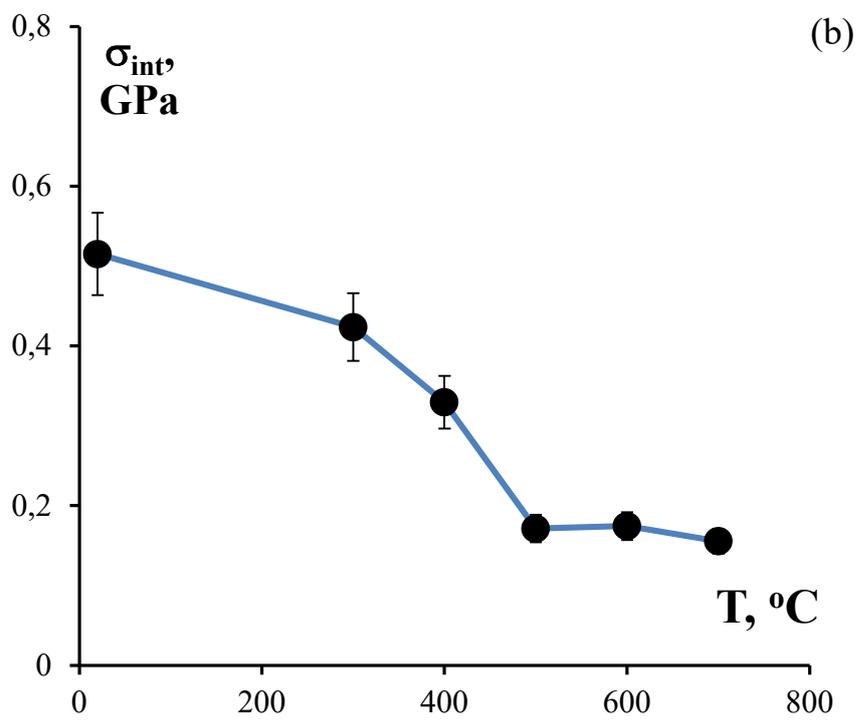

**Figure 12**

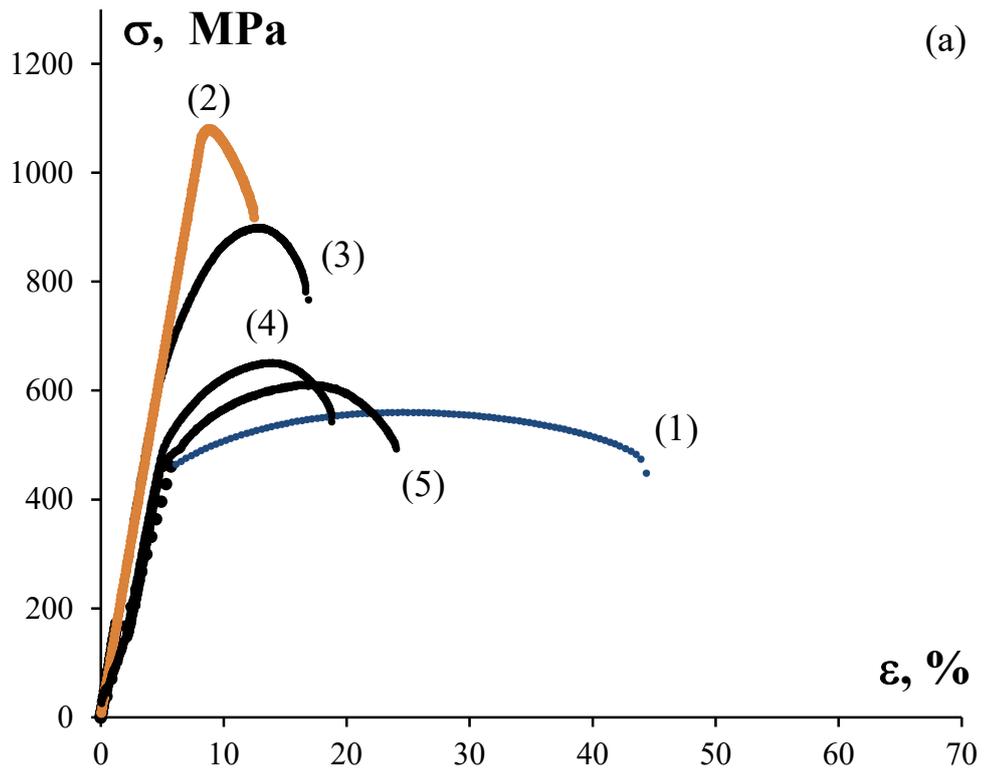

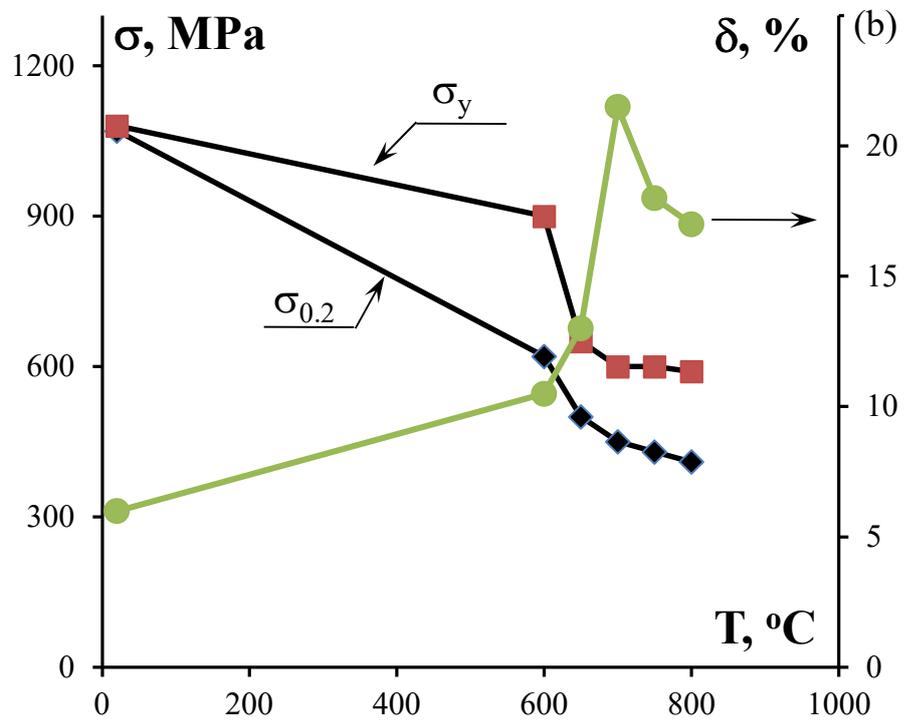

**Figure 13**

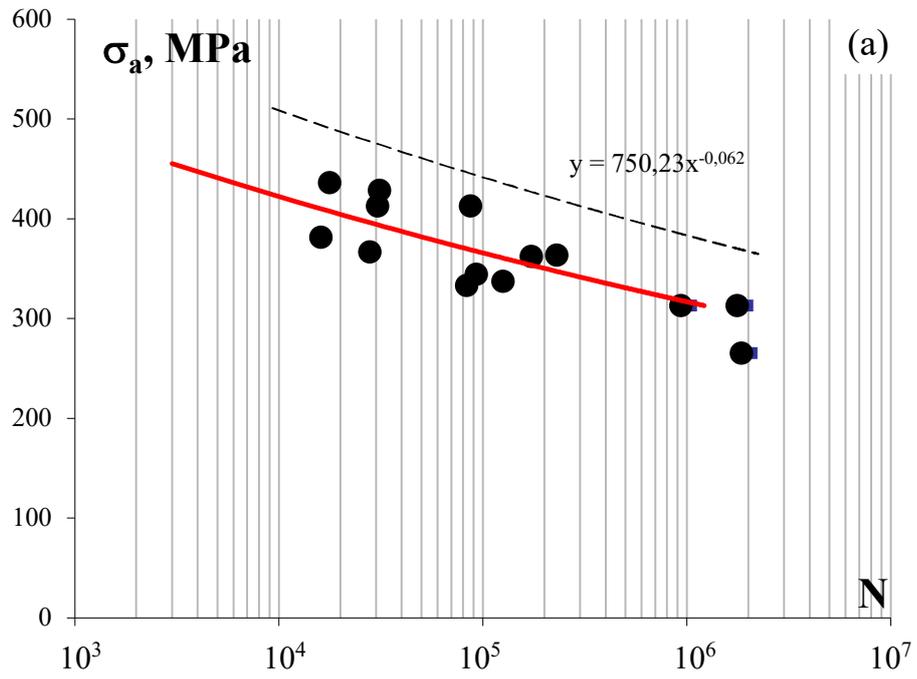

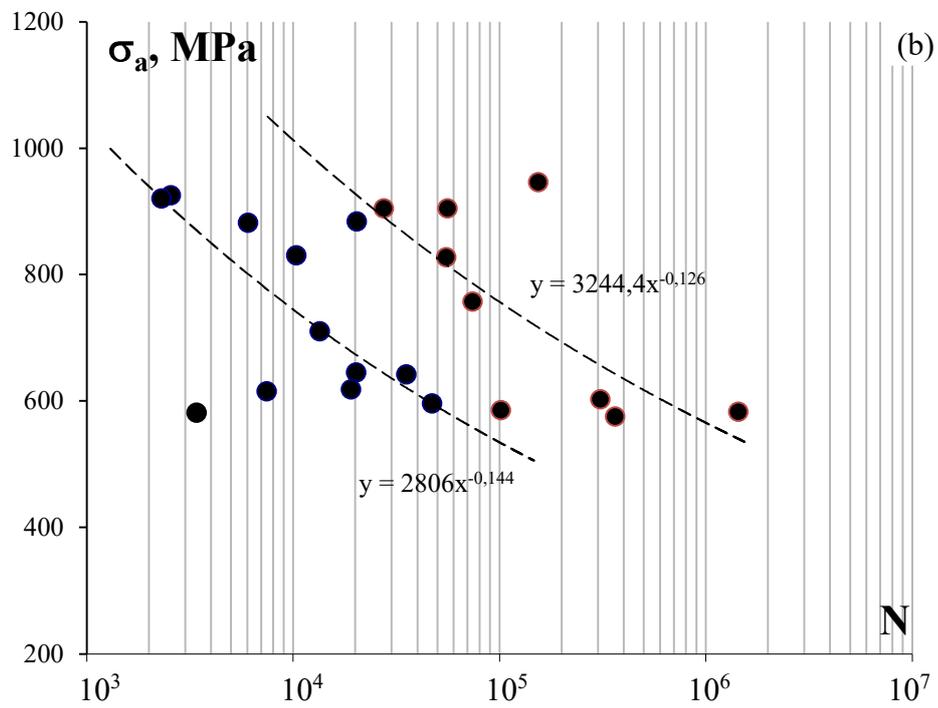

**Figure 14a, b**

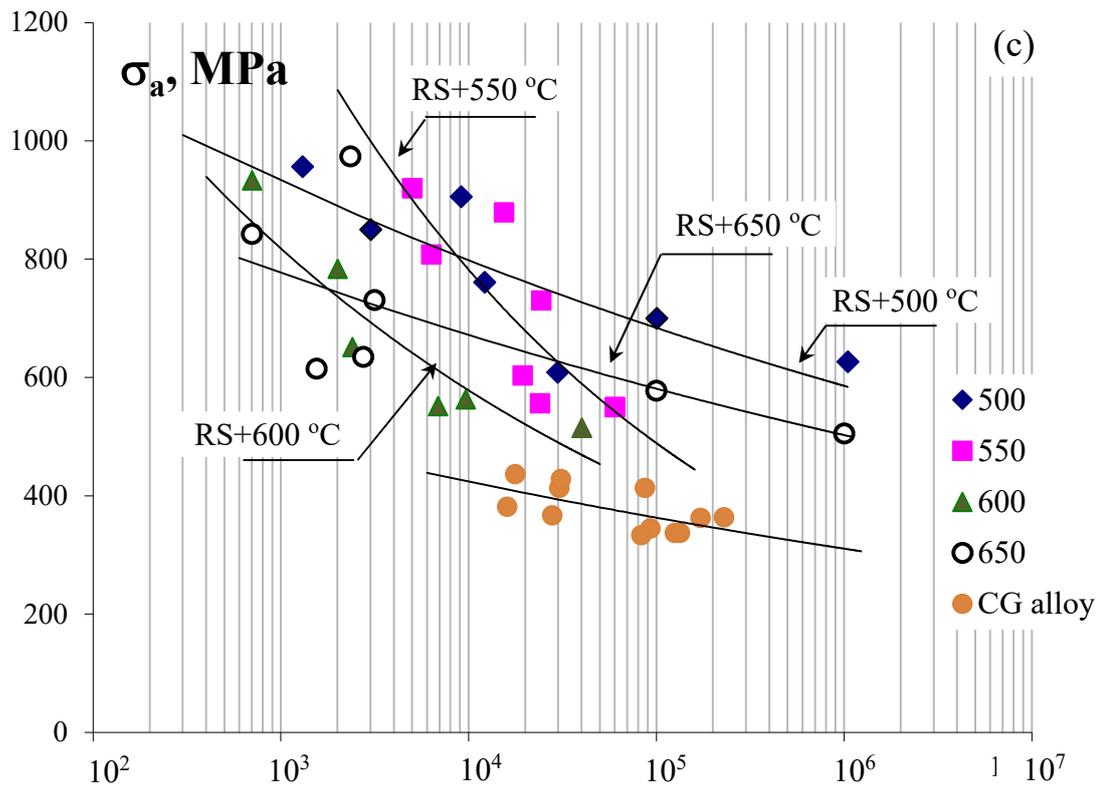

**Figure 14c**

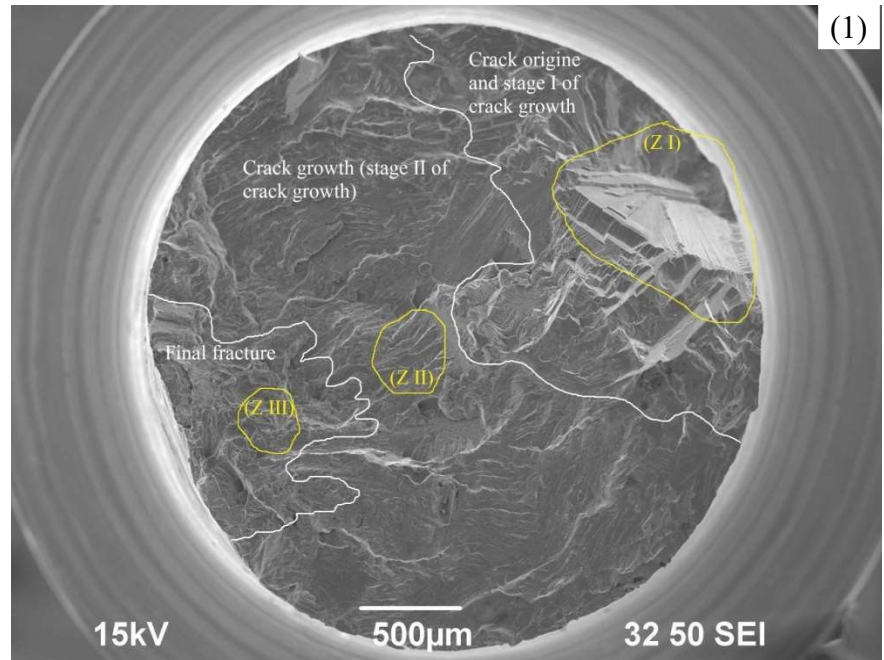
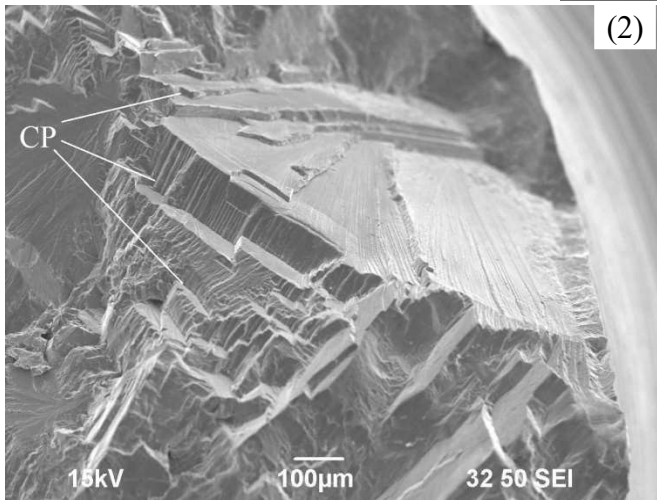
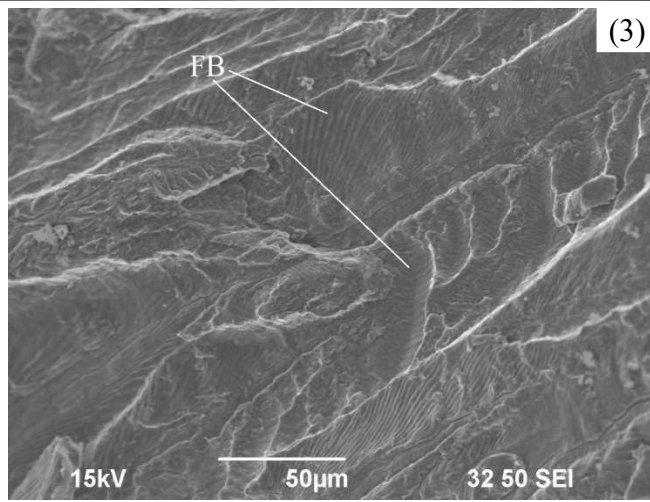
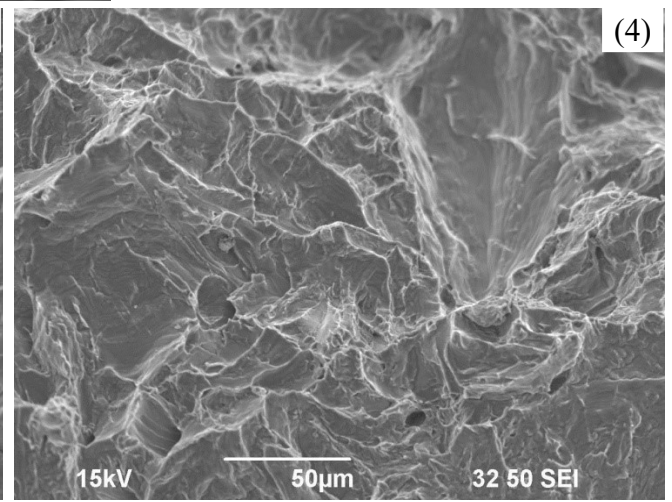

**Figure 15a**

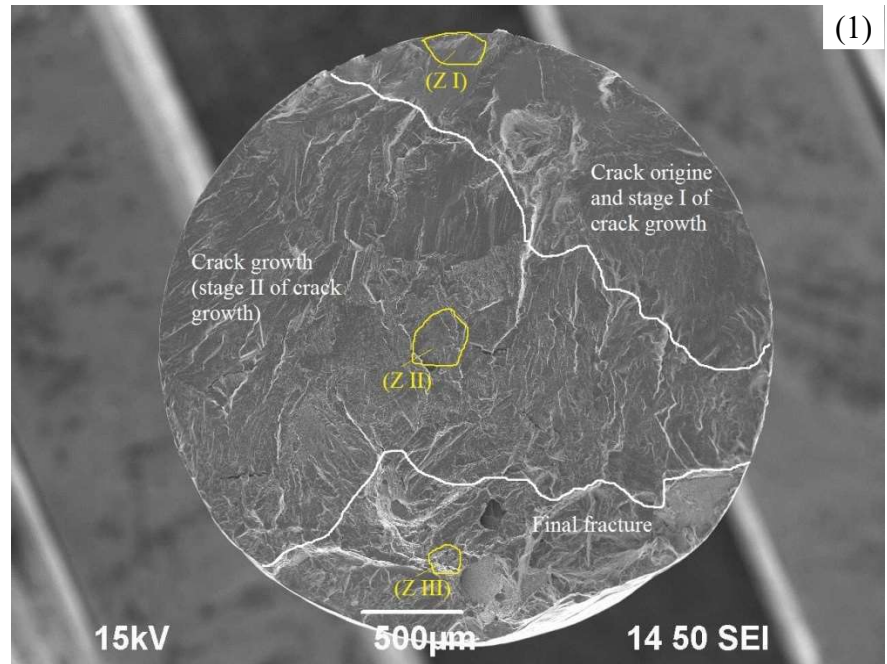
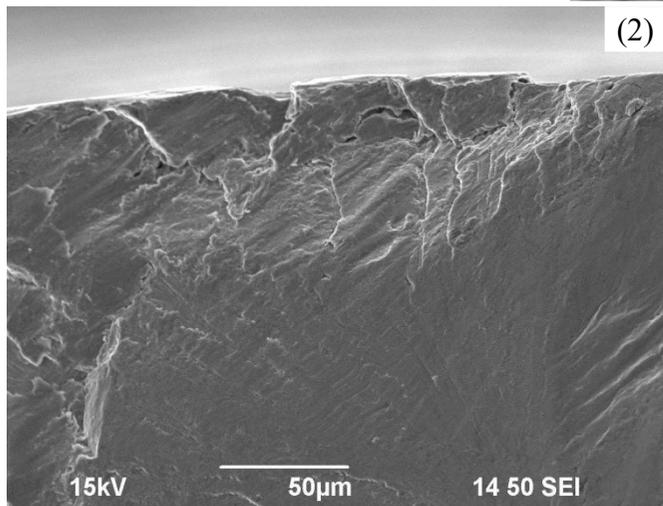 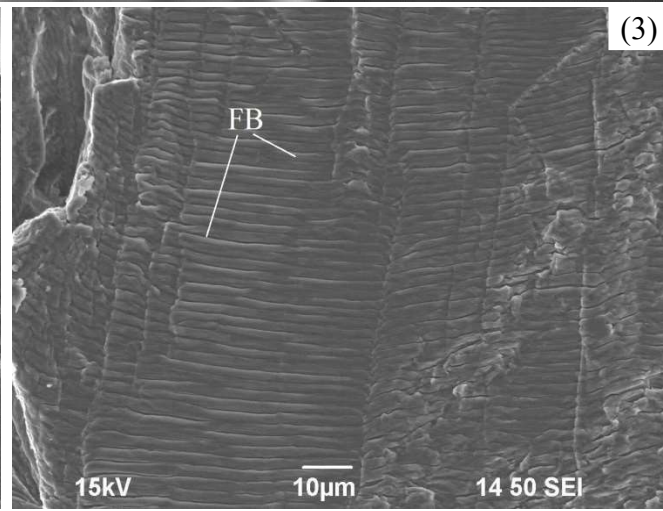 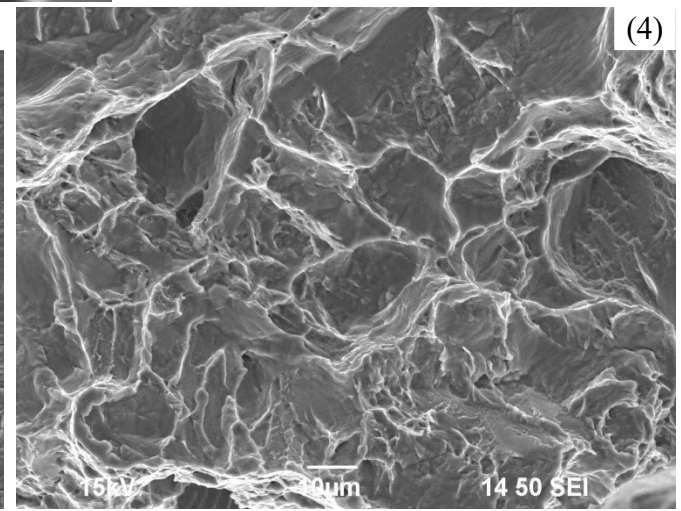

**Figure 15b**

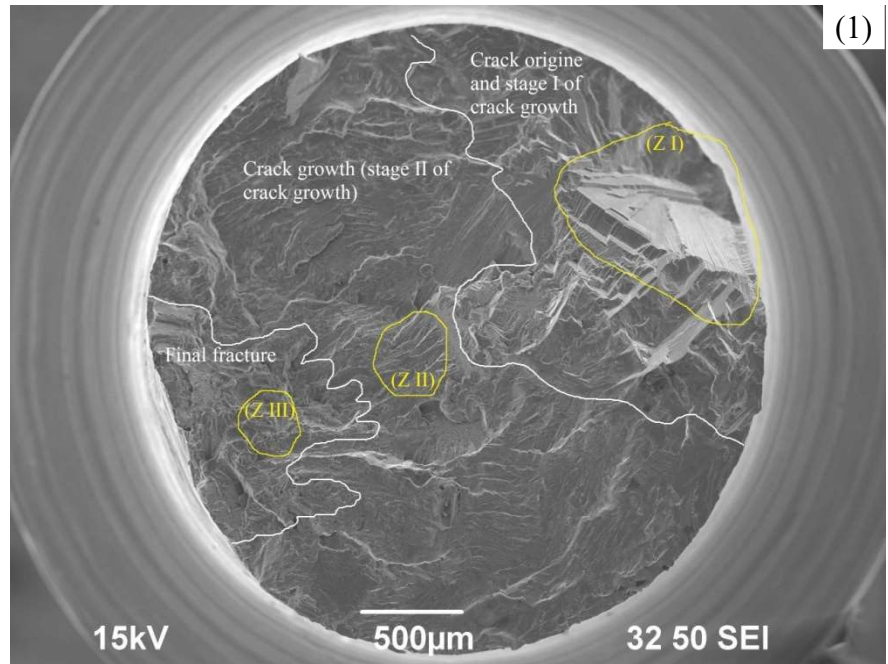
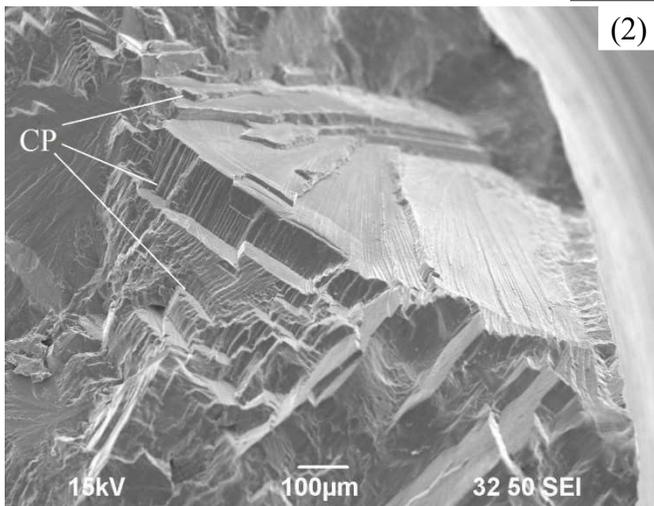
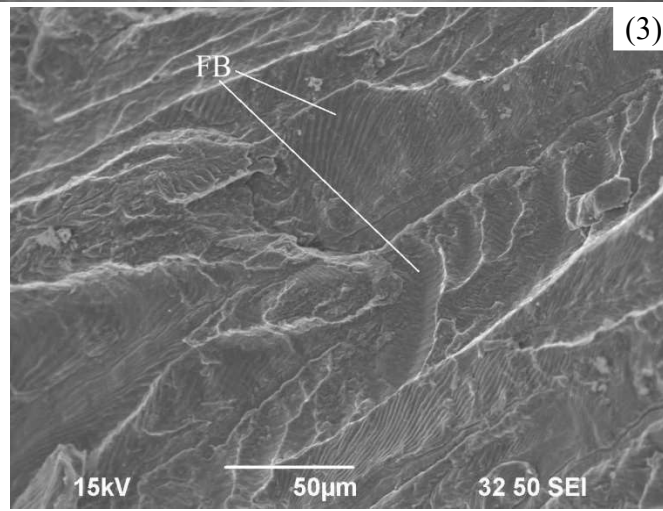
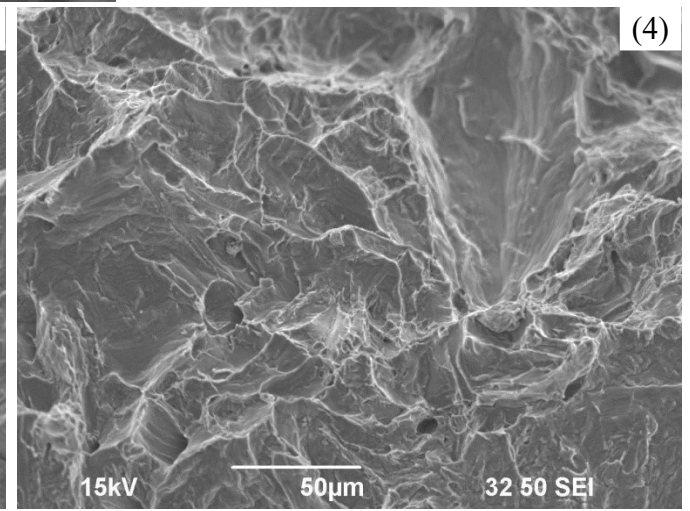

**Figure 15c**

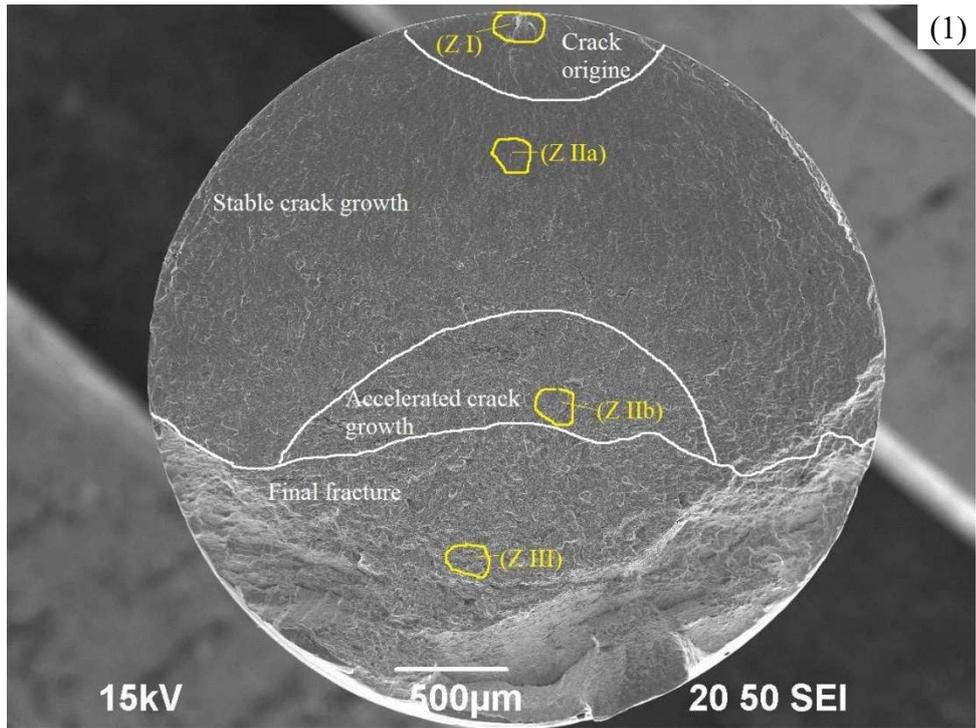
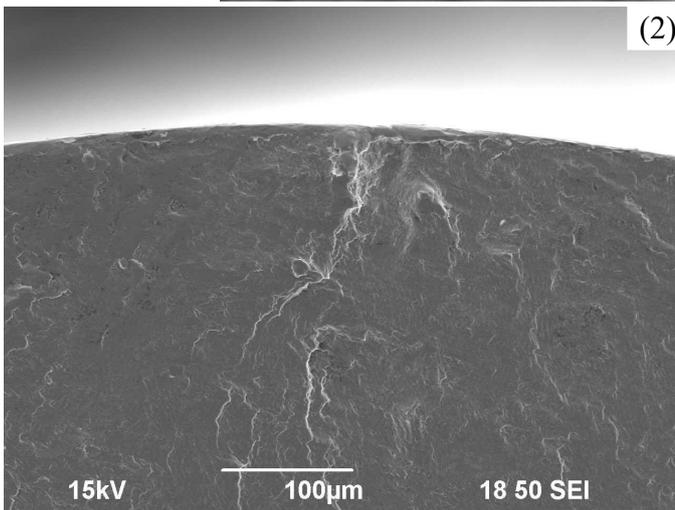
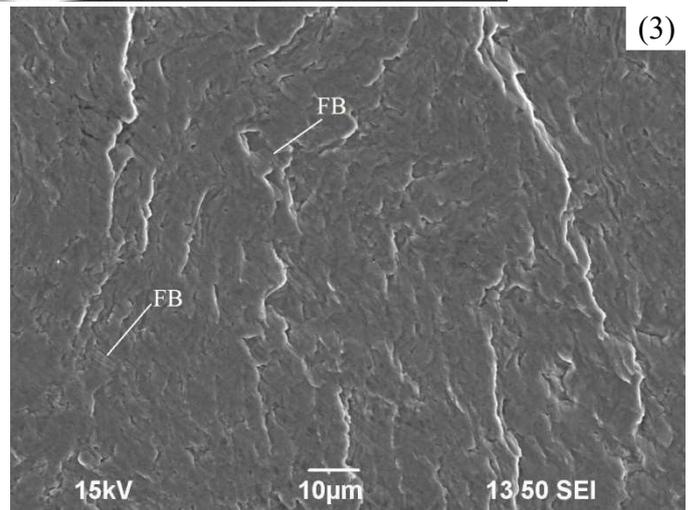
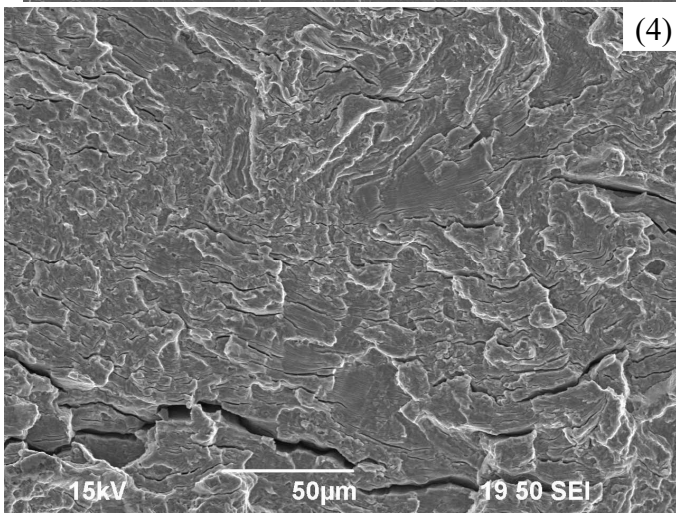
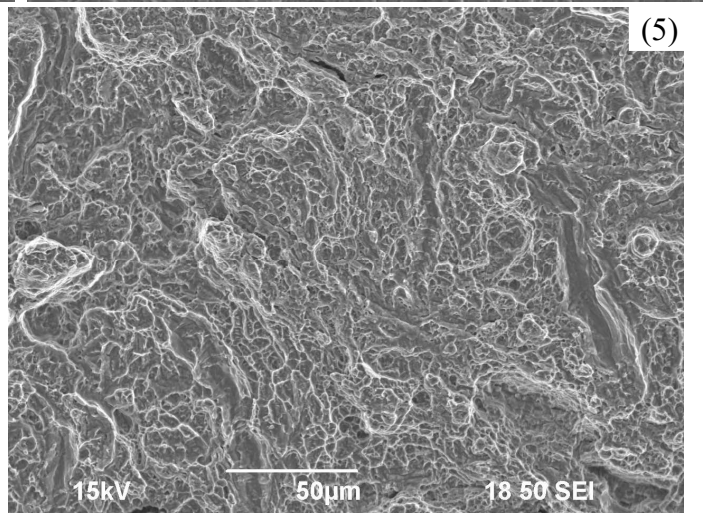

**Figure 16a**

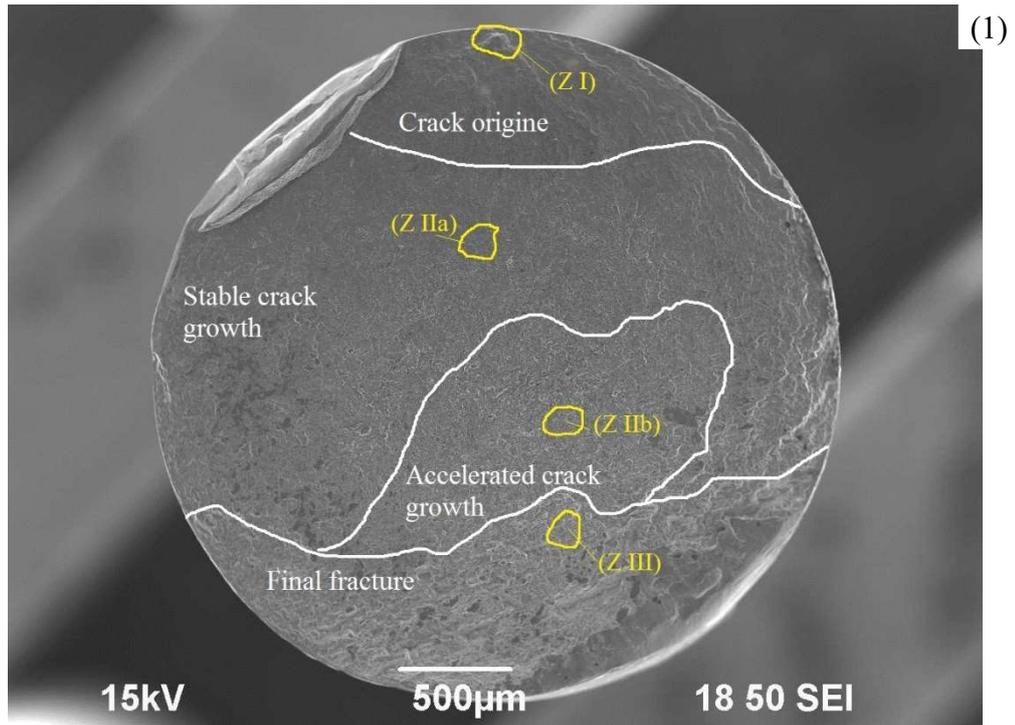

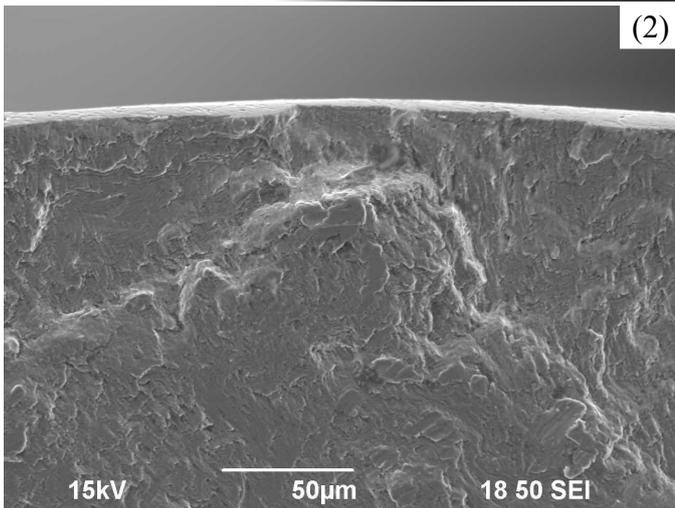
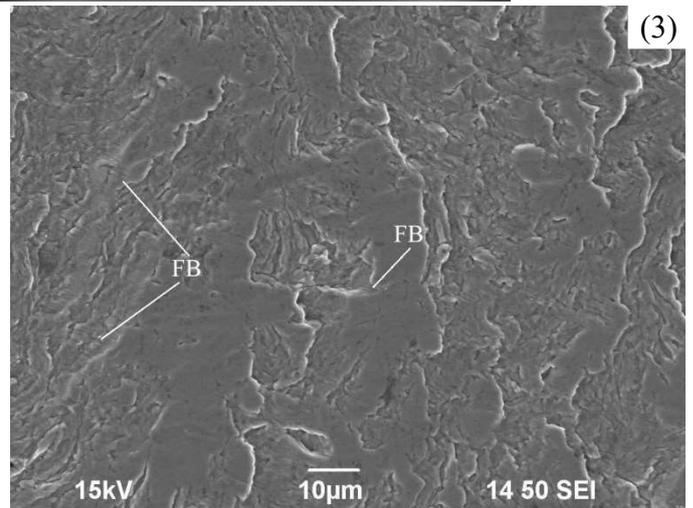

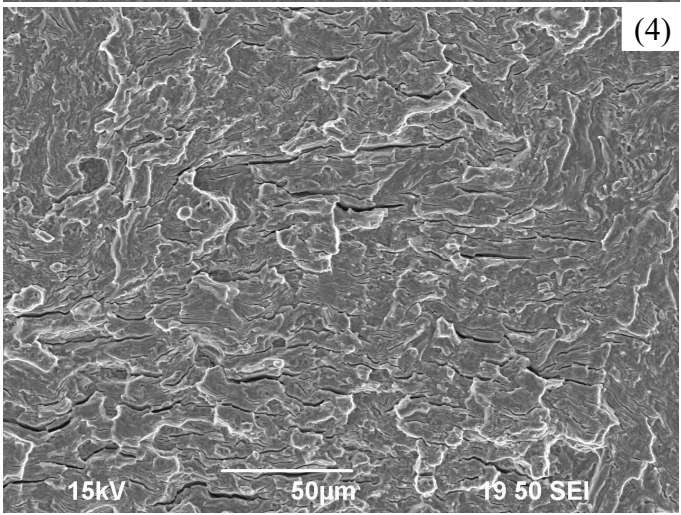
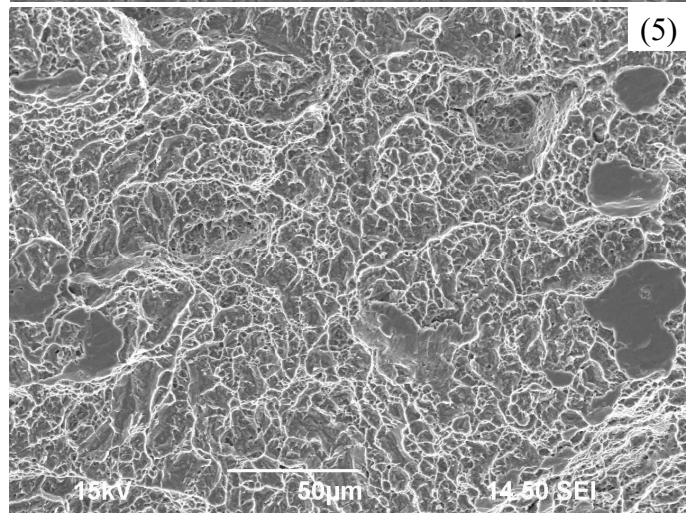

**Figure 16b**

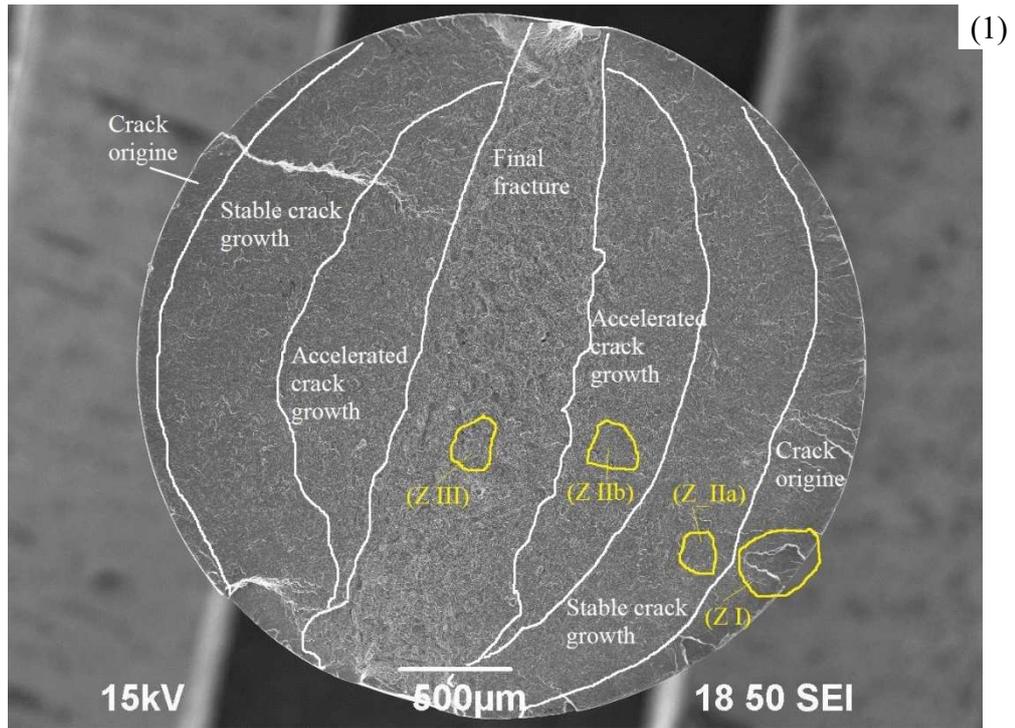
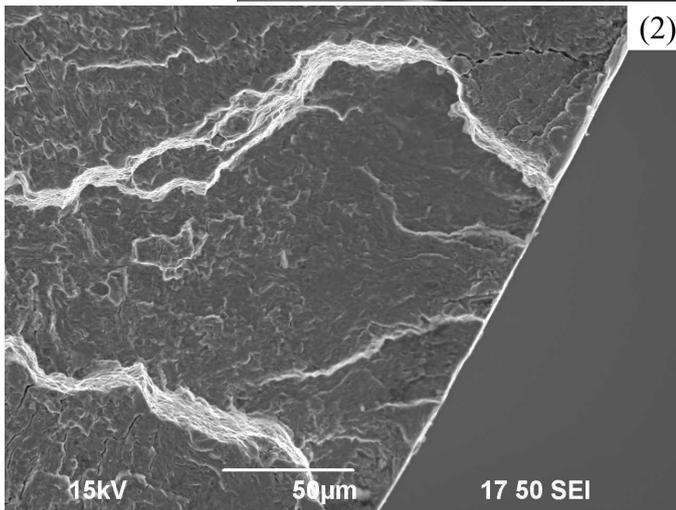
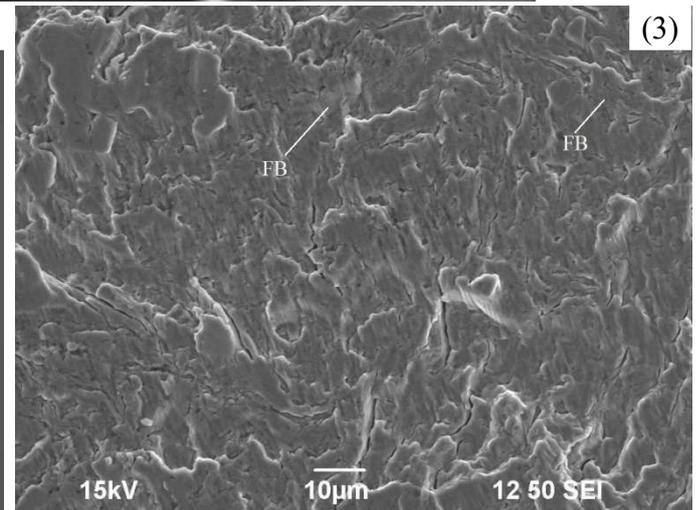
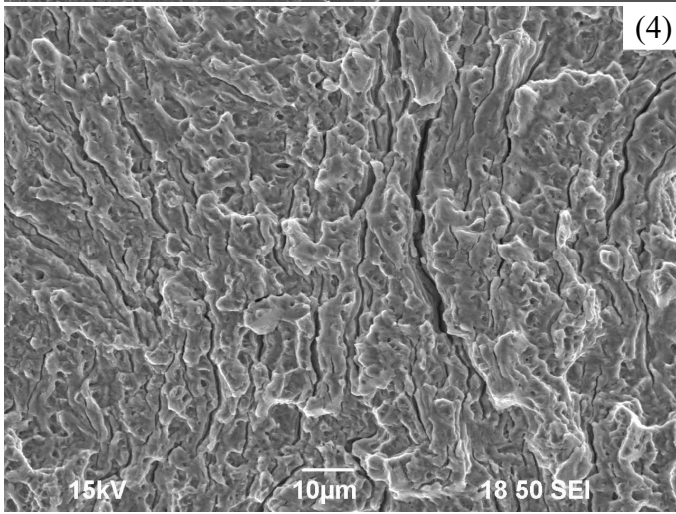
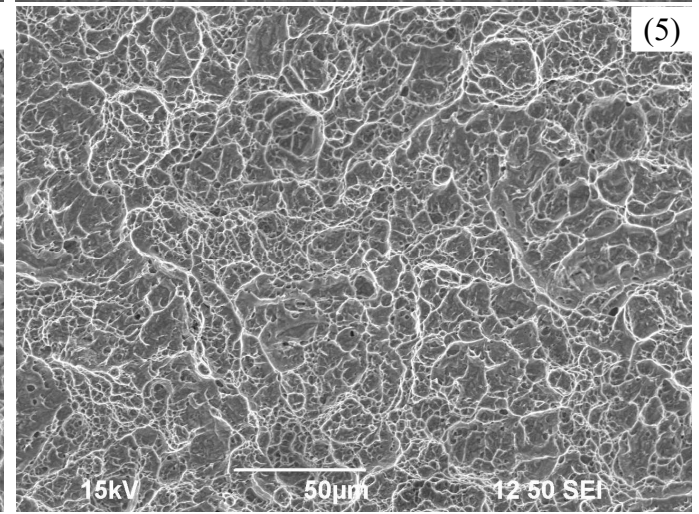

**Figure 16c**

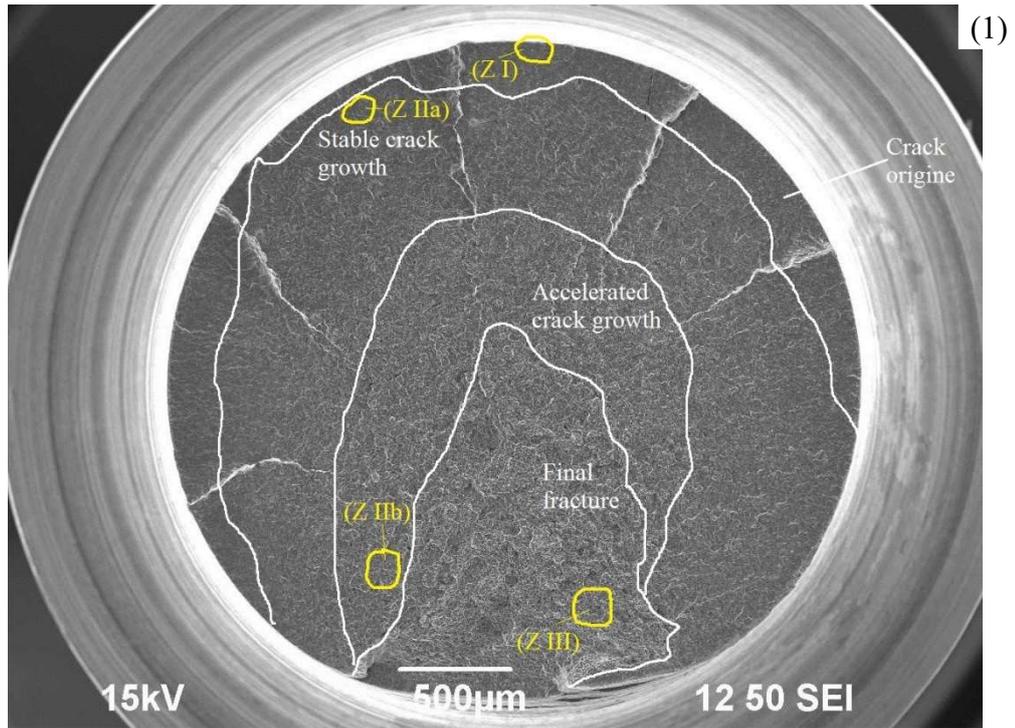

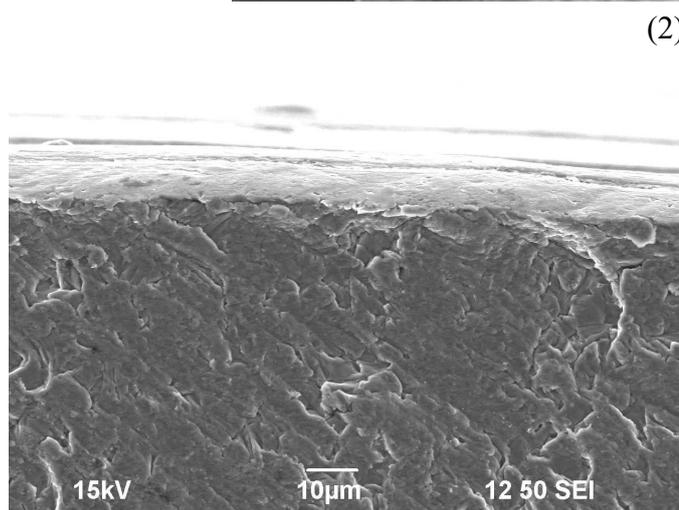
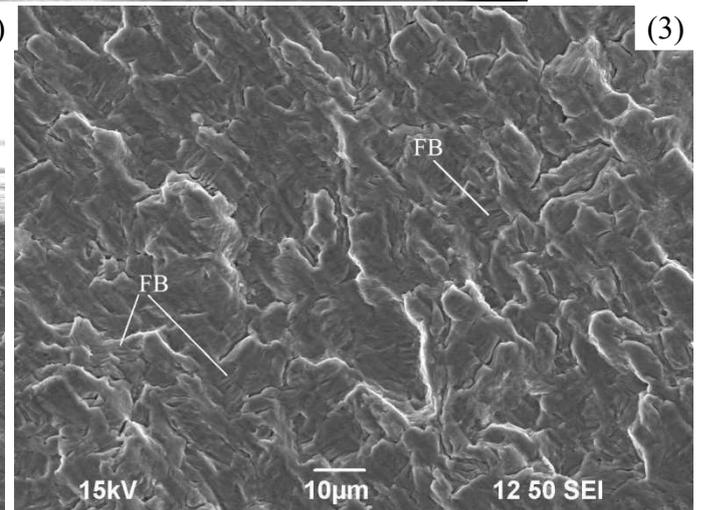
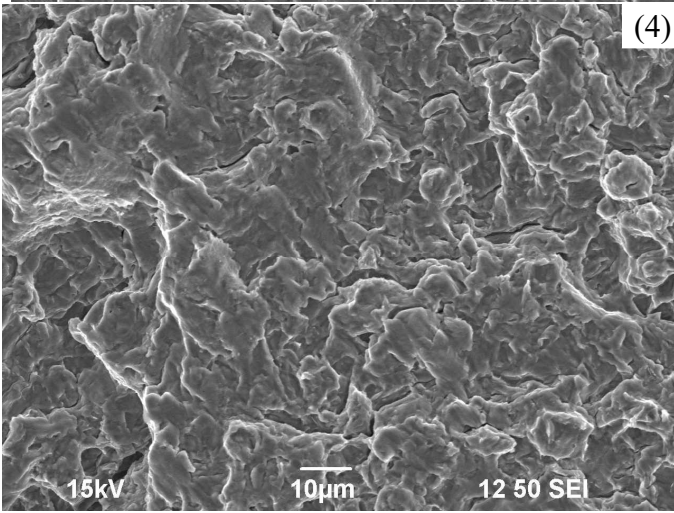
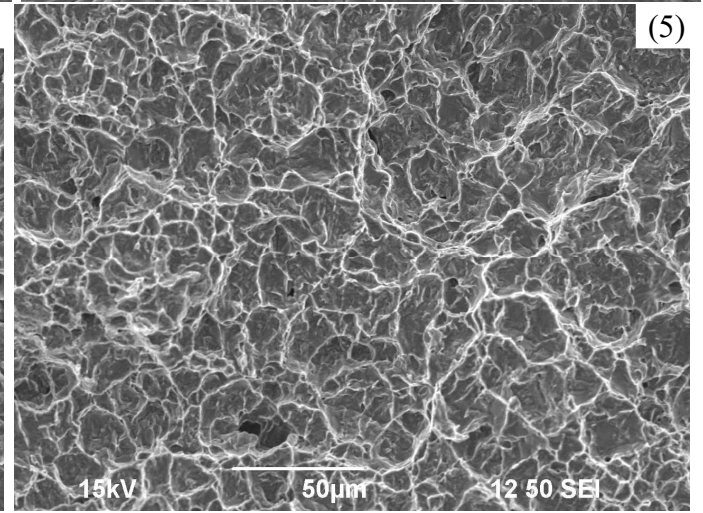

**Figure 17a**

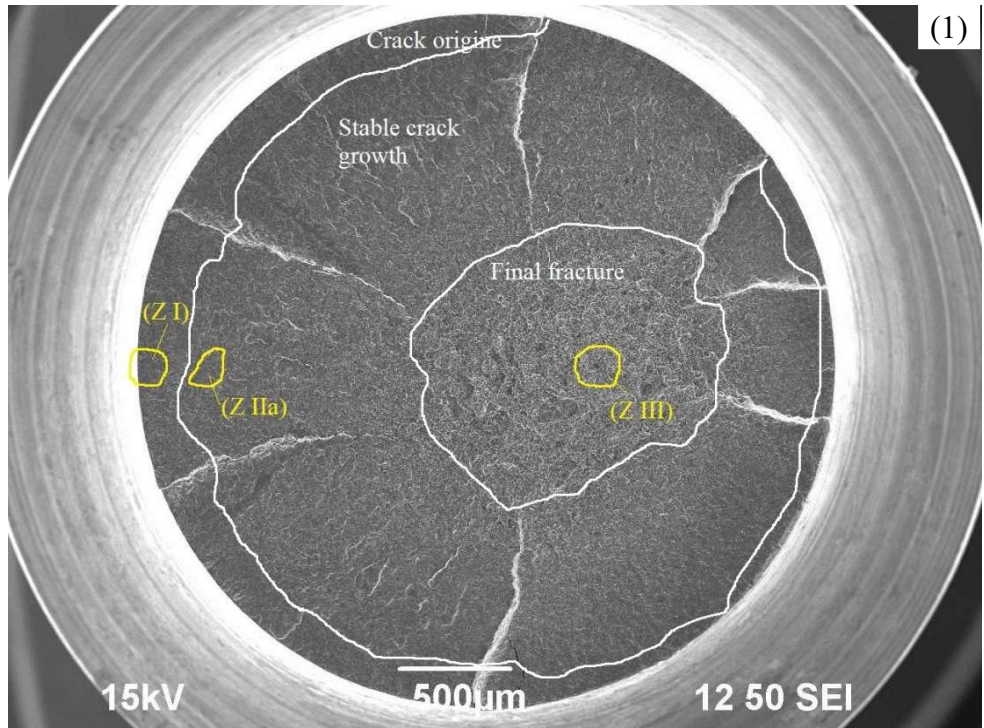
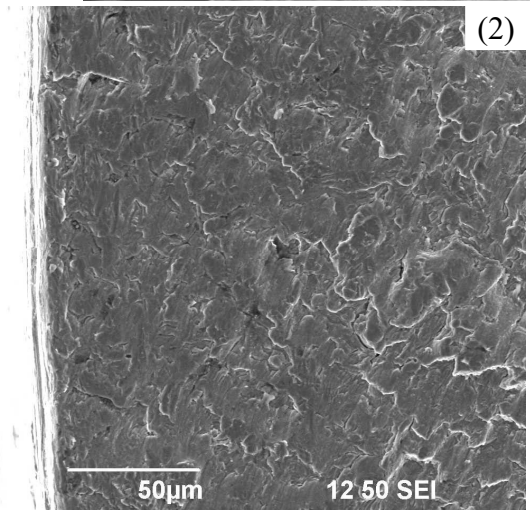
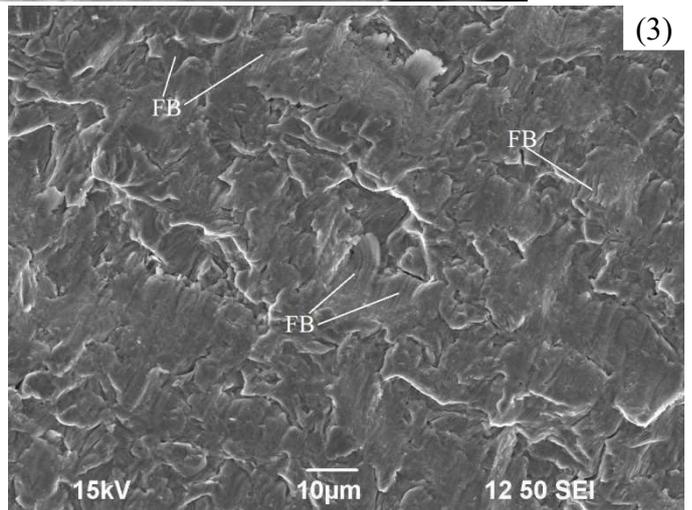
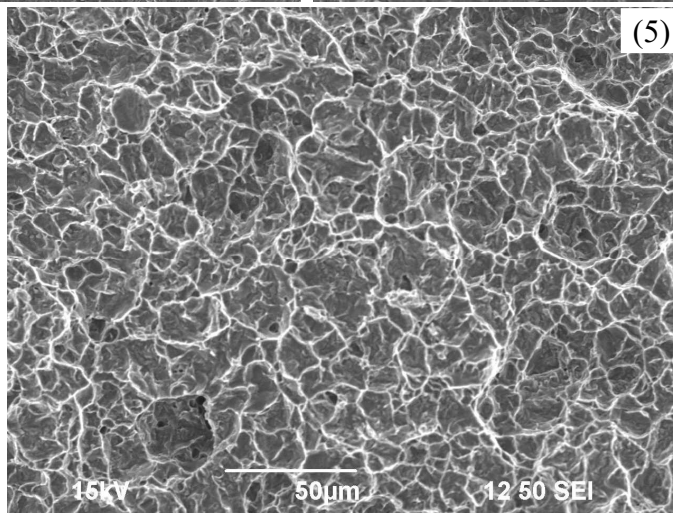

**Figure 17b**

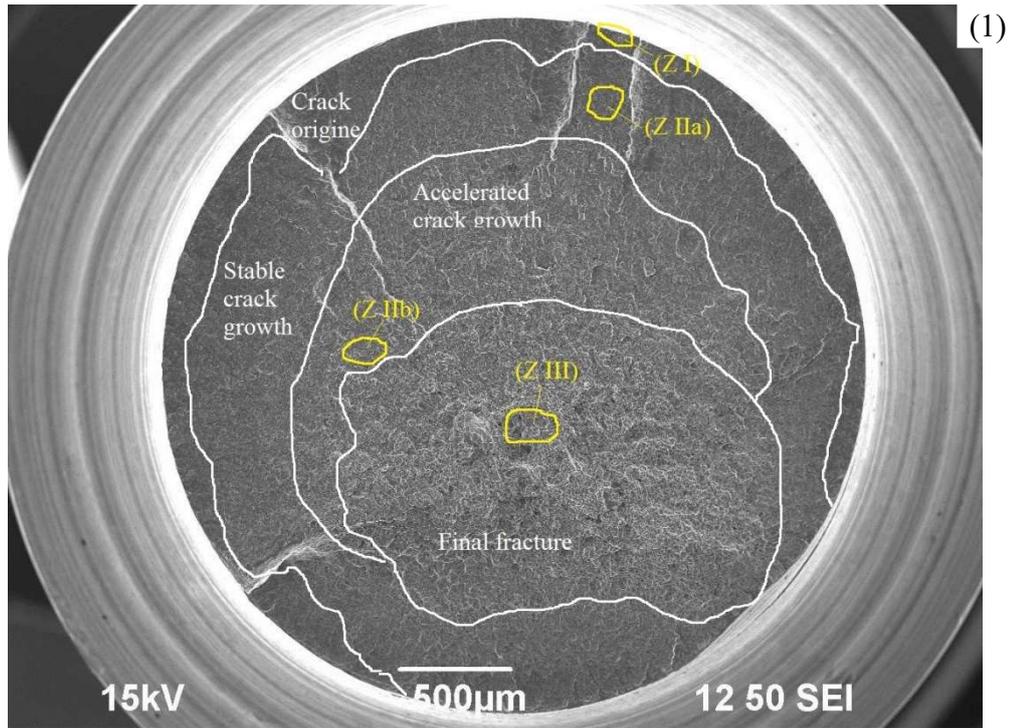
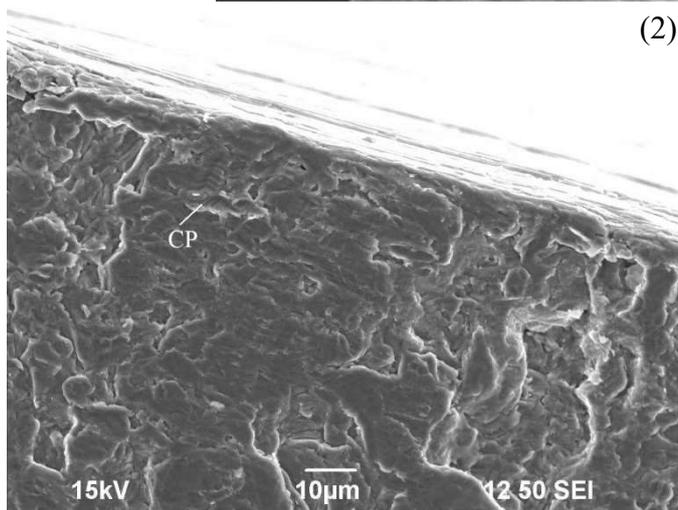
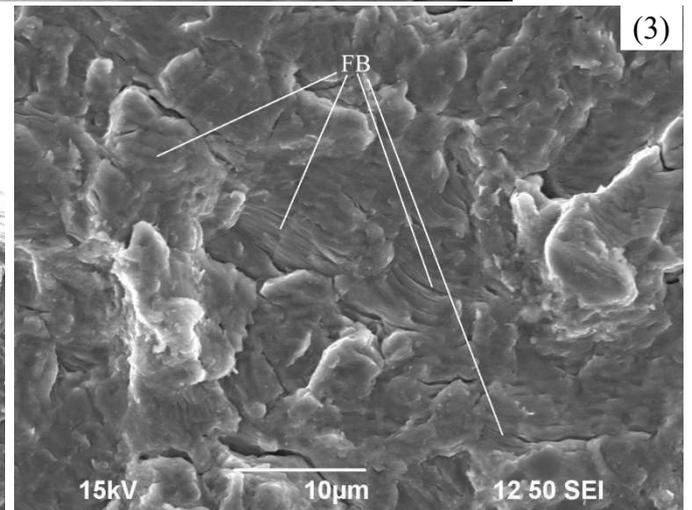
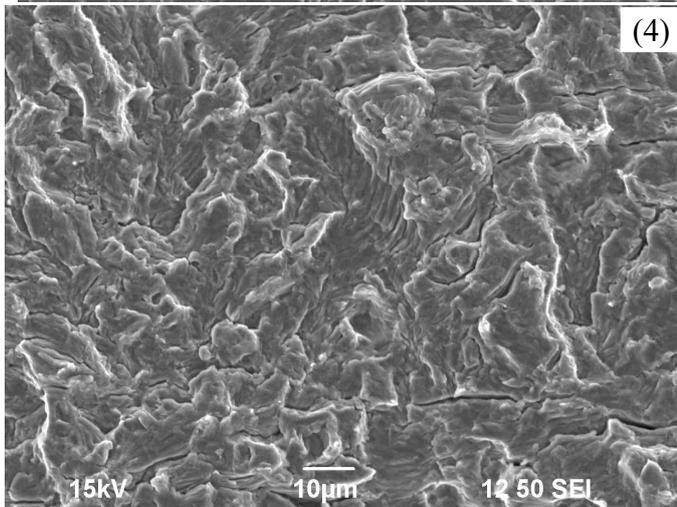
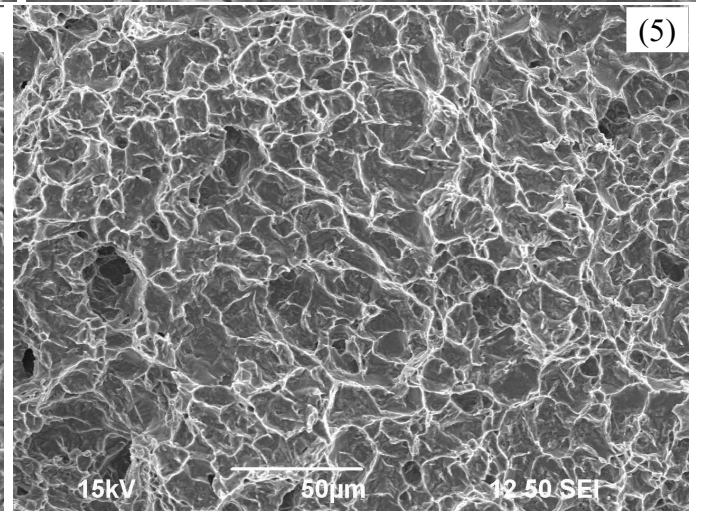

**Figure 17c**

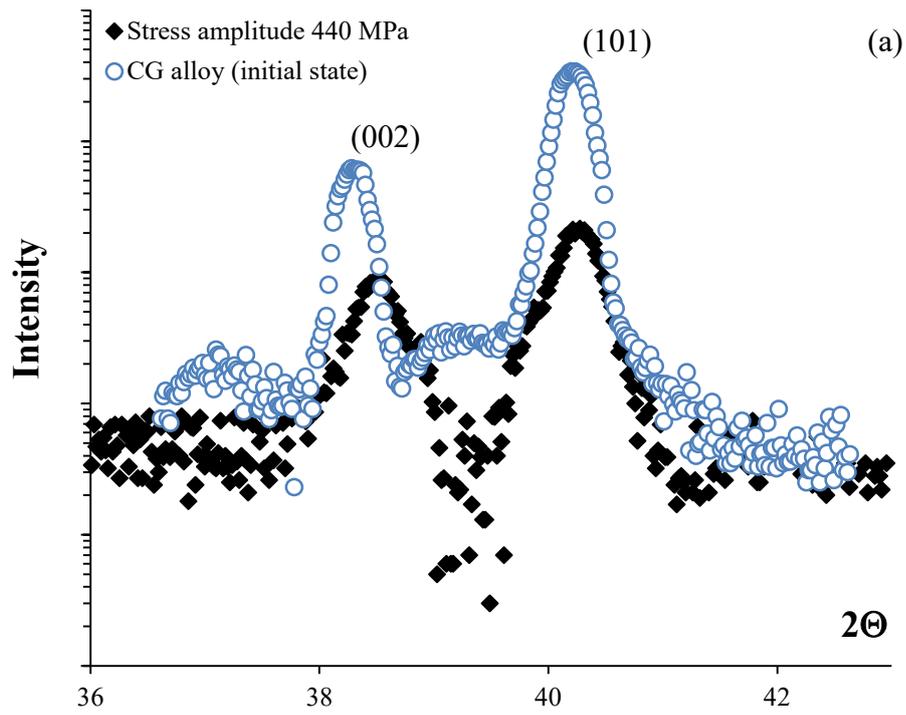

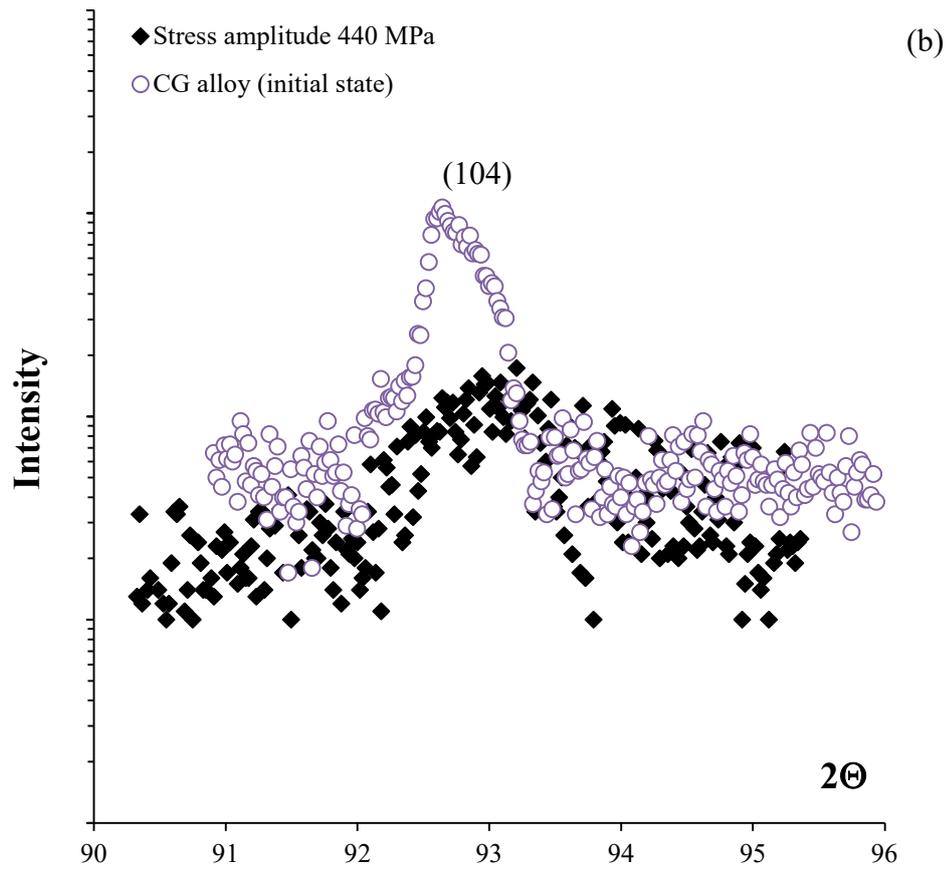

**Figure 18**

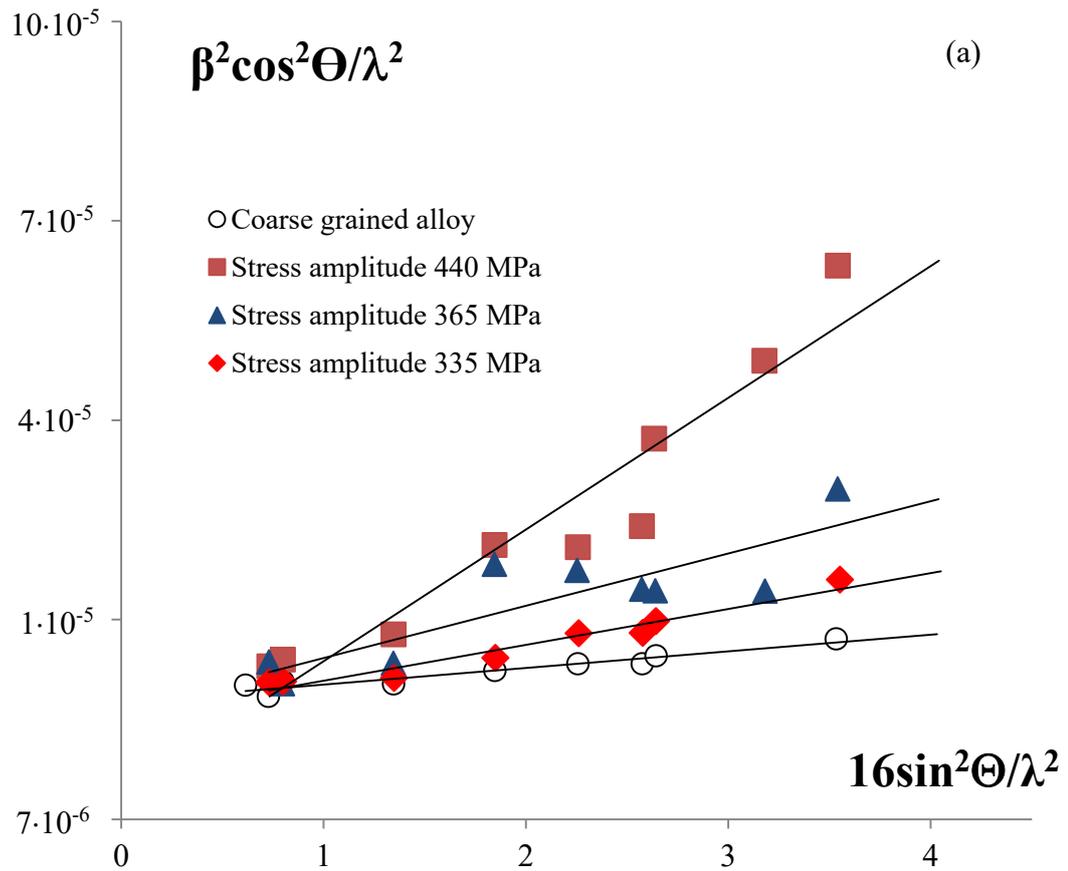

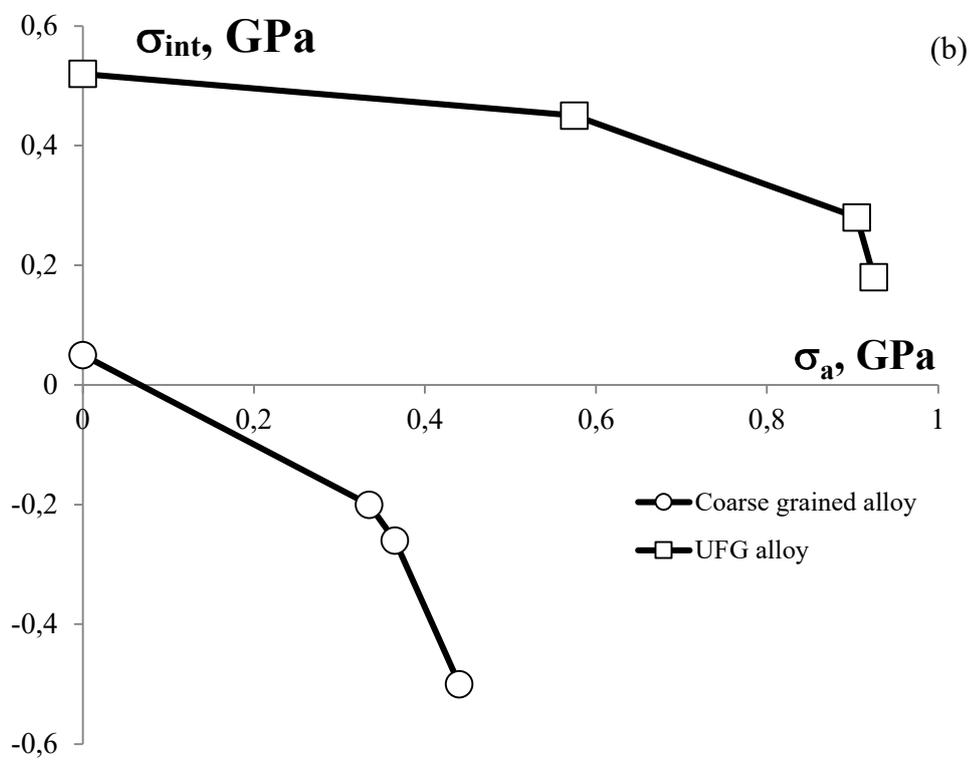

**Figure 19**